\documentclass{article}

\usepackage{arxiv}

\usepackage[T1]{fontenc}    
\usepackage{hyperref}       
\usepackage{url}            
\usepackage{booktabs}       
\usepackage{amsfonts}       
\usepackage{nicefrac}       
\usepackage{microtype}      
\usepackage{lipsum}
\usepackage{lineno,hyperref}
\usepackage{amsthm,amsmath,amssymb}
\usepackage[latin1,utf8]{inputenc}
\usepackage{algorithm}
\usepackage[noend]{algpseudocode}
\usepackage{etoolbox}
\usepackage{multirow}
\usepackage{pdflscape}
\usepackage{float}
\usepackage{graphicx}
\usepackage{xcolor}
\usepackage{caption}
\usepackage{verbatim}
\usepackage{geometry}
\usepackage{longtable}
\usepackage{subfig}

\usepackage{pifont}
\newcommand{\cmark}{\ding{51}}%
\newcommand{\xmark}{\ding{55}}%

\title{Branch-and-cut algorithms for the covering salesman problem}

\author{
  Lucas Porto Maziero \\
  Institute of Computing \\
  University of Campinas \\
  Av. Albert Einstein 1251, 13083-852, Campinas, SP, Brazil \\
  \texttt{lucas.maziero@ic.unicamp.br} \\
   \And
 Fábio Luiz Usberti \\
  Institute of Computing \\
  University of Campinas \\
  Av. Albert Einstein 1251, 13083-852, Campinas, SP, Brazil \\
  \texttt{fusberti@ic.unicamp.br} \\
  \And
  Celso Cavellucci \\
  Institute of Computing \\
  University of Campinas \\
  Av. Albert Einstein 1251, 13083-852, Campinas, SP, Brazil \\
  \texttt{celsocv@ic.unicamp.br} \\
}

\begin{document}
\maketitle

\begin{abstract}
The Covering Salesman Problem (CSP) is a generalization of the Traveling Salesman Problem in which the tour is not required to visit all vertices, as long as all vertices are covered by the tour. The objective of CSP is to find a minimum length Hamiltonian cycle over a subset of vertices that covers an undirected graph.
In this paper, valid inequalities from the generalized traveling salesman problem are applied to the CSP in addition to new valid inequalities that explore distinct aspects of the problem. A branch-and-cut framework assembles exact and heuristic separation routines for integer and fractional CSP solutions.
Computational experiments show that the proposed framework outperformed methodologies from literature with respect to optimality gaps. Moreover, optimal solutions were proven for several previously unsolved instances.
\end{abstract}

\keywords{Covering salesman problem \and integer linear programming \and branch-and-cut algorithm}

\section{Introduction} \label{sec:introduction}

Consider a set of sites scattered in the plane that must be covered by a single-vehicle tour. Knowing that each site covers some of its neighbors, what is the minimum length of an enclosed vehicle tour in which all sites are covered? This question is addressed by the Covering Salesman Problem (CSP), proposed by Current and Schilling \cite{current1989covering} in 1989. More formally, given an undirected graph, the CSP objective is to find the shortest Hamiltonian cycle on a subset of vertices that covers the graph.
The special case where each vertex covers strictly itself is the Travelling Salesman Problem (TSP) \cite{applegate2007}, which follows that CSP is also NP-hard.

Since its proposal, the CSP has attracted the attention of researchers due to its complexity and numerous applications. These applications arise in scenarios where it is unrealistic to visit all locations, e.g., rural health services, areas affected by natural disasters, or planning mobile service units \cite{current1989covering}.

Several variants of CSP were investigated. Current et al.~\cite{current1994efficient} studied the Shortest Covering Path Problem (SCPP). The goal is to find a minimum cost $s$-$t$ path in a network that covers all vertices. The authors proposed two methods to solve the SCPP: a Lagrangian relaxation and a branch-and-bound algorithm that makes use of the obtained dual bounds.

Current and Schilling~\cite{current1994median} introduced two bi-criterion routing problems: the Median Tour Problem (MTP) and the Maximal Covering Tour Problem (MCTP). Assuming a network with $n$ vertices and a value $p$ $(p \leqslant n)$, the criteria for both problems are $(i)$ to find a minimum length tour that visits exactly $p$ of the $n$ vertices and $(ii)$ to maximize the accessibility of the vertices that are not in the tour. The problems differ in the way the accessibility of the second criterion is evaluated. In MTP, the second criterion is to minimize the sum of distances from each unvisited vertex to its closest vertex in the tour. In MCTP, the second criterion is to minimize the number of uncovered vertices. The authors proposed mathematical formulations and heuristics to solve both MTP and MCTP. Their methodologies were tested on a real-life scenario requiring the optimal location and sequence of stops for overnight mail service.

Another variant of CSP, studied by Gendreau et al. \cite{gendreau1997covering}, is the Covering Tour Problem (CTP). Let $G = (V \cup W, E)$ be an undirected graph, where $V \cup W$ is the set of vertices and $E$ is the set of edges. Vertex $v_0$ is the depot, $V$ is the set of vertices that can be visited, $T \subseteq V$ is the set of vertices that must be visited $(v_0 \in T)$, and $ W $ is the set of vertices that must be covered but cannot be visited. The goal of the CTP is to determine a minimum length tour that visits a subset of vertices $S \subseteq V$ such that $T \subseteq S$ and each vertex of $W$ is covered by some vertex in $S$. The authors proposed heuristics and a branch-and-cut algorithm to solve the CTP.

Golden et al.~\cite{golden2012generalized} proposed a generalized version of the CSP called the Generalized Covering Salesman Problem (GCSP). Given an undirected graph $G = (V, E)$, each vertex $i \in V$ has a covering demand $k_i$, meaning vertex $i$ has to be covered at least $k_i$ times. In addition, there is a fixed cost $F_{i}$ that incurs when the tour visits vertex $i$.  The objective of the GCSP is to minimize the solution cost which is given by the sum of the tour length and the costs of the visited vertices. The authors developed local searches that explore exchange, removal, and insertion of tour vertices to escape from local optimum.

Another similar problem to the CSP is the Generalized Traveling Salesman Problem (GTSP). In GTSP, the vertices are partitioned into disjoint subsets, called clusters, and the goal is to determine the minimum length tour that visits exactly one vertex from each cluster. The GTSP is a special case of the CSP, where each cluster can be modeled as a subset of vertices that mutually cover themselves. Fischetti, González, and Toth~\cite{fischetti1997branch} propose a branch-and-cut algorithm based on exact and heuristic separation routines for some families of valid inequalities for the GTSP.
These inequalities are translated for the CSP in Section~\ref{sec:validInequality}.

Zhang and Xu~\cite{zhang2018online} propose the online CSP, where the vehicle will face up to $k$ blocked edges not known a priori during its tour traversal. The objective is to find a minimum length tour that covers all vertices while bypassing the blocked edges. The authors presented a $(k + \alpha)$-competitive algorithm, where $\alpha = \frac{1}{2} + \frac{(4k + 2)r}{OPT} + 2\upsilon\rho$, $\upsilon$ is the approximation ratio for the Steiner Tree Problem, $\rho$ is the maximum number of vertices that can cover an arbitrary vertex and $r$ is the radius which defines the covering neighbourhood of each vertex.

Many works in literature have given attention to the geometric CSP, also known as the Close Enough Traveling Salesman Problem (CETSP). In this version, each vertex has its neighborhood defined as a compact region of the plane. The goal is to find a minimum length tour that starts from a depot and intercepts all neighborhood sets, thus covering all its corresponding vertices. Approximation algorithms, heuristics and methodologies based on ILP were developed for this version (Dumitrescu and Mitchell~\cite{dumitrescu2003approximation}, Gulczynski et al.~\cite{gulczynski2006close}, Dong et al.~\cite{dong2007heuristic}, Shuttleworth et al.~\cite{shuttleworth2008advances}, Behdani and Smith~\cite{behdani2014integer}, Coutinho et al.~\cite{coutinho2016branch}).

Table~\ref{tab:coveringProblemComparison} emphasizes the main differences between CSP and its counterparts. In CTP, among the vertices that can be visited, for some of them the visitation is mandatory. As for the vertices that must be covered, in CTP these vertices cannot be visited. In GTSP, the vertices are clustered into disjoint neighborhoods, meaning each vertex covers exactly the vertices in the cluster it belongs. The vertices in GCSP may require multiple coverings and each visitation incurs into a fixed cost. Finally, in CETSP the vertices are covered by a compact region on the plane instead of being covered by a subset of vertices. All of these problems, despite sharing the idea of joining vehicle routing with set covering, contain important distinctions with respect to CSP. This explains why this problem still requires customized exact and heuristic methodologies.

\begin{table}[]
\setlength{\tabcolsep}{3pt}
\scriptsize
\caption{Summary of the main differences between CSP, CTP, GTSP, GCSP and CETSP.}
\label{tab:coveringProblemComparison}
\centering
\begin{tabular}{lcccc}
               & \multicolumn{1}{c}{\textbf{\begin{tabular}[c]{@{}c@{}}Required/Forbidden\\ visitations\end{tabular}}} & \multicolumn{1}{c}{\textbf{\begin{tabular}[c]{@{}c@{}}Disjoint\\ neighborhoods\end{tabular}}} & \multicolumn{1}{c}{\textbf{\begin{tabular}[c]{@{}c@{}}Multiple\\ coverings\end{tabular}}} &  \multicolumn{1}{c}{\textbf{\begin{tabular}[c]{@{}c@{}}Geometric\\ neighborhood\end{tabular}}} \\ \hline
\textbf{CSP}   & \xmark                                                                                                                                 & \xmark                                                                                                                                                             & \xmark                                                                                                                                                                                                                        & \xmark                                                                                                                        \\
\textbf{CTP}   & \cmark                                                                                                                                 & \xmark                                                                                                                                                             & \xmark                                                                                                                                                                                                                            & \xmark                                                                                                                        \\
\textbf{GTSP}  & \xmark                                                                                                                                 & \cmark                                                                                                                                                             & \xmark                                                                                                                                                                                                                             & \xmark                                                                                                                        \\
\textbf{GCSP}  & \xmark                                                                                                                                 & \xmark                                                                                                                                                             & \cmark                                                                                                                                                                                                                             & \xmark                                                                                                                        \\
\textbf{CETSP} & \xmark                                                                                                                                 & \xmark                                                                                                                                                             & \xmark                                                                                                                                                                                                                             & \cmark                                                                                                                        \\ \hline
\end{tabular}
\end{table}

Some solution methodologies were proposed in the literature for the CSP. Current and Schilling \cite{current1989covering}, for example, developed a two-step heuristic to solve the CSP: the first step solves a set cover problem; the second step solves the TSP on the vertices determined by the first step. More than two decades later Salari and Naju-Azimi~\cite{salari2012integer} revisited the problem by proposing a heuristic for the CSP embedded within an integer linear programming (ILP) framework. First, they employ constructive heuristics to find good initial solutions and then the tour vertices are rearranged by the use of ILP techniques in an attempt to reduce its length. Salari et al.~\cite{salari2015combining} give a polynomial-size formulation and a hybrid heuristic for the CSP, which combines ant colony optimization and dynamic programming. The formulation of Salari et al., to the best of our knowledge, composes the state-of-the-art exact methodology for the CSP.

More recently, Venkatesh et al.~\cite{venkatesh2019multi} proposed a multi-start iterated local search algorithm for the CSP and incorporated a variable degree of perturbation strategy to further improve the solution obtained through their heuristic approach. Computational results show that the proposed approach is competitive with other state-of-the-art heuristic approaches for solving the CSP. Zang et al.~\cite{zang2020parallel} reformulated the CSP as a bilevel CSP (BCSP) with a leader and a follower sub-problem and proposed two parallel variable neighborhood search (PVNS) heuristics, namely, synchronous “master–slave” PVNS and asynchronous cooperative PVNS. Computational results show that the PVNS has improved previously best known solutions. Venkatesh et al.~\cite{pandiri2020two} developed two hybrid metaheuristic approaches for the CSP. The first is based on the artificial bee colony algorithm (ABC) and the second is based on the genetic algorithm (GA). Both approaches were competitive with the state-of-the-art of heuristic approaches for the CSP.


\textbf{Our contribution:} despite being well studied in the heuristic front, the CSP still lacks effective exact methods. Many of the current best known solutions still had not been proven optimal or had an open optimality gap due to the absence of a dual bound. To address this matter, the first branch-and-cut framework is proposed for the CSP. The framework employs exact and heuristic routines to separate families of valid inequalities, some from the GTSP and others original for the CSP. Computational experiments performed on a benchmark set of instances compares our methodology with the state-of-the-art exact methodology from literature. Previous to this work, from a set of $48$ instances, for only $9$ instances there were proven optimal solutions. Our methodology  improves this by certifying optimality for all except one instance. This represents a major  contribution to the current body of knowledge regarding exact approaches on the CSP.

This paper is organized as follows. Section~\ref{sec:csp} formally defines the CSP and presents an integer linear programming formulation. Section~\ref{sec:validInequality} shows new valid inequalities for the CSP. Section~\ref{sec:cutSeparation} describes separation routines for the proposed valid inequalities, which constitute the branch-and-cut framework. In Section~\ref{sec:experiments} computational experiments are conducted on a benchmark of instances, and results are analyzed and discussed. Section~\ref{sec:conclusions} gives the concluding remarks.

\section{Problem Description and Formulation} \label{sec:csp}

\begin{figure*}[htbp]
  \centering
  \subfloat[Instance kroB200-7]{	
		\includegraphics[width=0.65\textwidth]{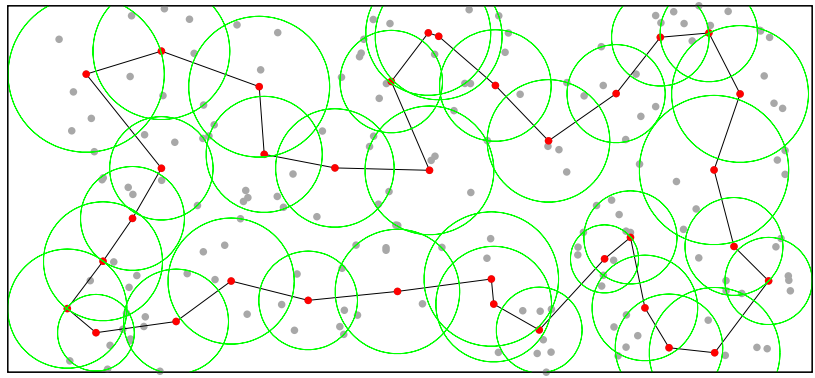}
		\label{fig:solutionCSPkroB200-7} }\\
  \subfloat[Instance kroB200-9]{	
		\includegraphics[width=0.65\textwidth]{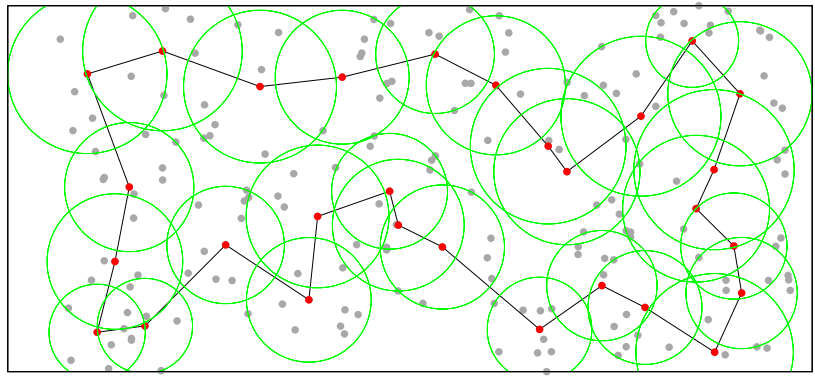}
		\label{fig:solutionCSPkroB200-9} }\\
  \subfloat[Instance kroB200-11]{	
		\includegraphics[width=0.65\textwidth]{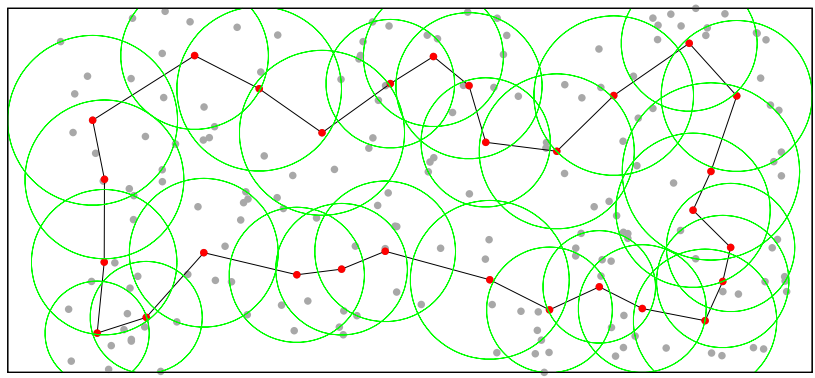}
		\label{fig:solutionCSPkroB200-11} }
	\caption{Optimal solutions for instances kroB200-7, kroB200-9, and kroB200-11, where each vertex covers its closest $7$, $9$, and $11$ neighbors, respectively. Highlighted vertices belong to the tour and their covering sets are represented by circumferences.}
	\label{fig:solutionCSPkroB200}
\end{figure*}

The CSP can be formally stated as follows. Consider an undirected graph $G(V, E)$, where $V$ is the set of vertices and $E$ is the set of edges. Each edge $e \in E$ is associated with a non-negative cost $c_e$. For each vertex $v \in V$, $C(v)$ is the set of vertices that cover $v$ and $D(v)$ is the set of vertices that are covered by $v$. It is considered that $v \in C(v)$ and $v \in D(v)$, $\forall v \in V$. An optimal solution to the CSP is a minimum length Hamiltonian cycle (tour) over a subset of vertices that covers all vertices in $G$. Figures \ref{fig:solutionCSPkroB200-7}, \ref{fig:solutionCSPkroB200-9} and \ref{fig:solutionCSPkroB200-11} show optimal solutions for three CSP instances with $200$ vertices.



An integer linear programming (ILP) formulation for the CSP is presented. Binary variable $x_e$ shows if an edge $e \in E$ belongs ($1$) or not ($0$) to the tour and binary variable $y_v$ shows if a vertex belongs ($1$) or not ($0$) to the tour. We denote $\delta(v)$ the set of edges incident to $v \in V$, $\delta(S)$ the set of edges with one endpoint in $S \subset V$ and the other in $V \backslash S$ and $E(S)$ the set of edges with both endpoints in $S$.

\begin{align} \label{eq:modeloCSP}
	& (CSP) \nonumber \\
	& MIN \quad \sum_{e \in E}{c_e x_e}, \\
	& \mbox{subject to} \nonumber \\
	& \sum_{e \in \delta(v)}{x_e} = 2y_v \qquad & \forall v \in V, \\
	& \sum_{i \in C(v)}{y_i} \geqslant 1 \qquad & \forall v \in V, \\
	& \sum_{e \in \delta(S)}{x_e} \geqslant 2(y_i + y_j - 1) \qquad & \forall S \subset V, i \in S, j \in V \backslash S, \label{eq:sec} \\	
	& x_{e} \in \{0,1\} \qquad & \forall i,j \in V, \\
	& y_v \in \{0,1\} \qquad & \forall i \in V.
\end{align}


The CSP formulation is derived from the GTSP formulation proposed by Fischetti, González, and Toth \cite{fischetti1997branch}. The objective function ($1$) minimizes the cost of a solution given by the sum of the costs of its edges. Constraints ($2$) ensure the number of edges incident at a vertex is $2$ (if $v$ is in the tour) or $0$ (otherwise). Constraints ($3$) impose that each vertex must be covered at least once. Constraints ($4$) are subtour elimination constraints which state that every cut separating two vertices in the tour contains at least two edges.

\section{Valid Inequalities} \label{sec:validInequality}

This section presents valid inequalities proposed by Fischetti, González, and Toth~\cite{fischetti1997branch} for the GTSP, and here translated for the CSP. It is worth reminding that the GTSP is a special case of the CSP in which the vertices are partitioned into clusters, and each cluster is formed by vertices which mutually cover themselves, i.e., any two vertices $u$ and $v$ from the same cluster would have $C(u) = C(v)$.

\begin{figure*}[htbp]
  \centering
  \subfloat[$S \notin \gamma(V)$]{	
		\includegraphics[width=0.35\textwidth]{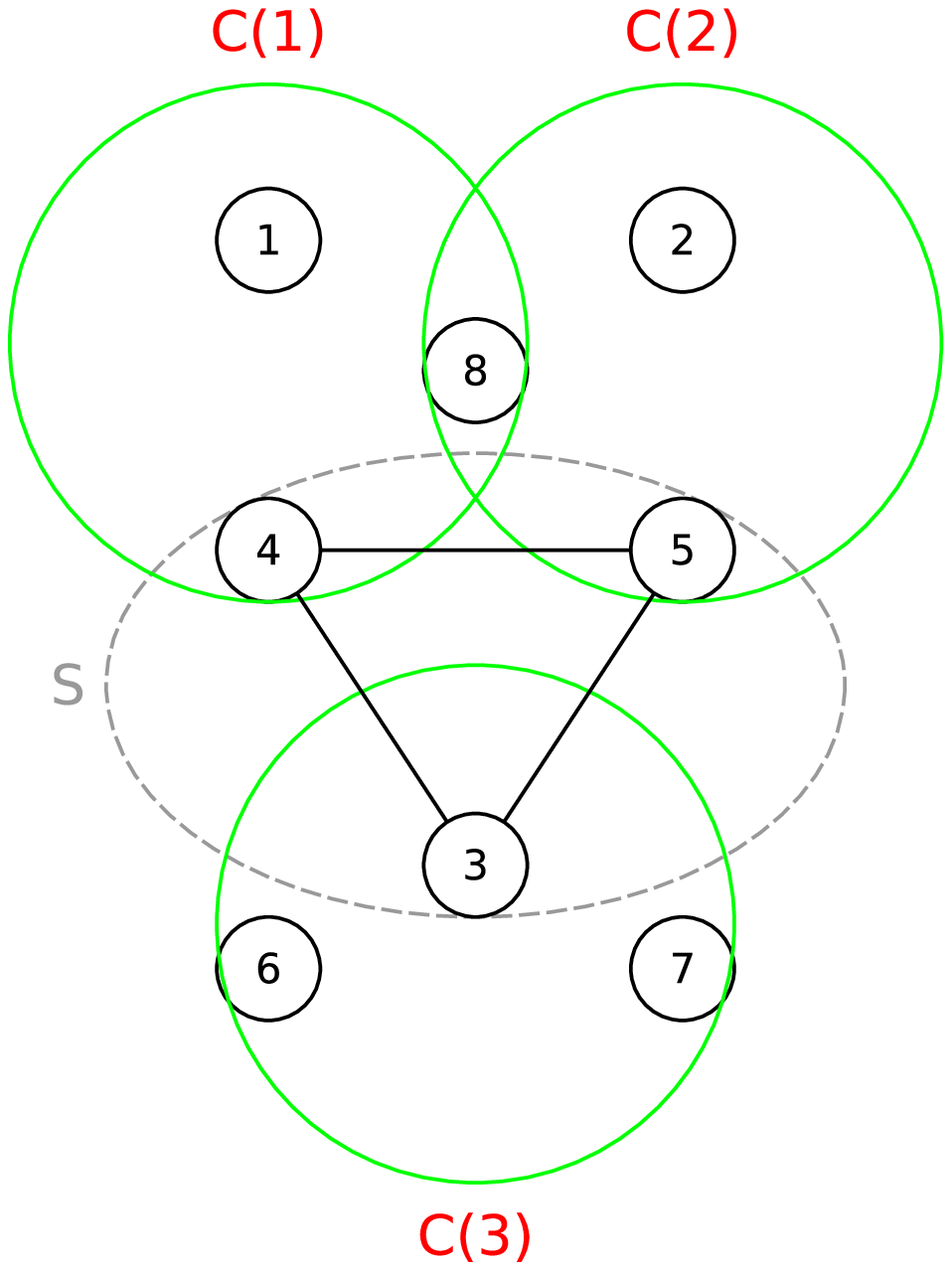}
		\label{fig:invalidGammaV} }
		\hspace{0.5cm}
  \subfloat[$S \in \gamma(V)$]{	
		\includegraphics[width=0.35\textwidth]{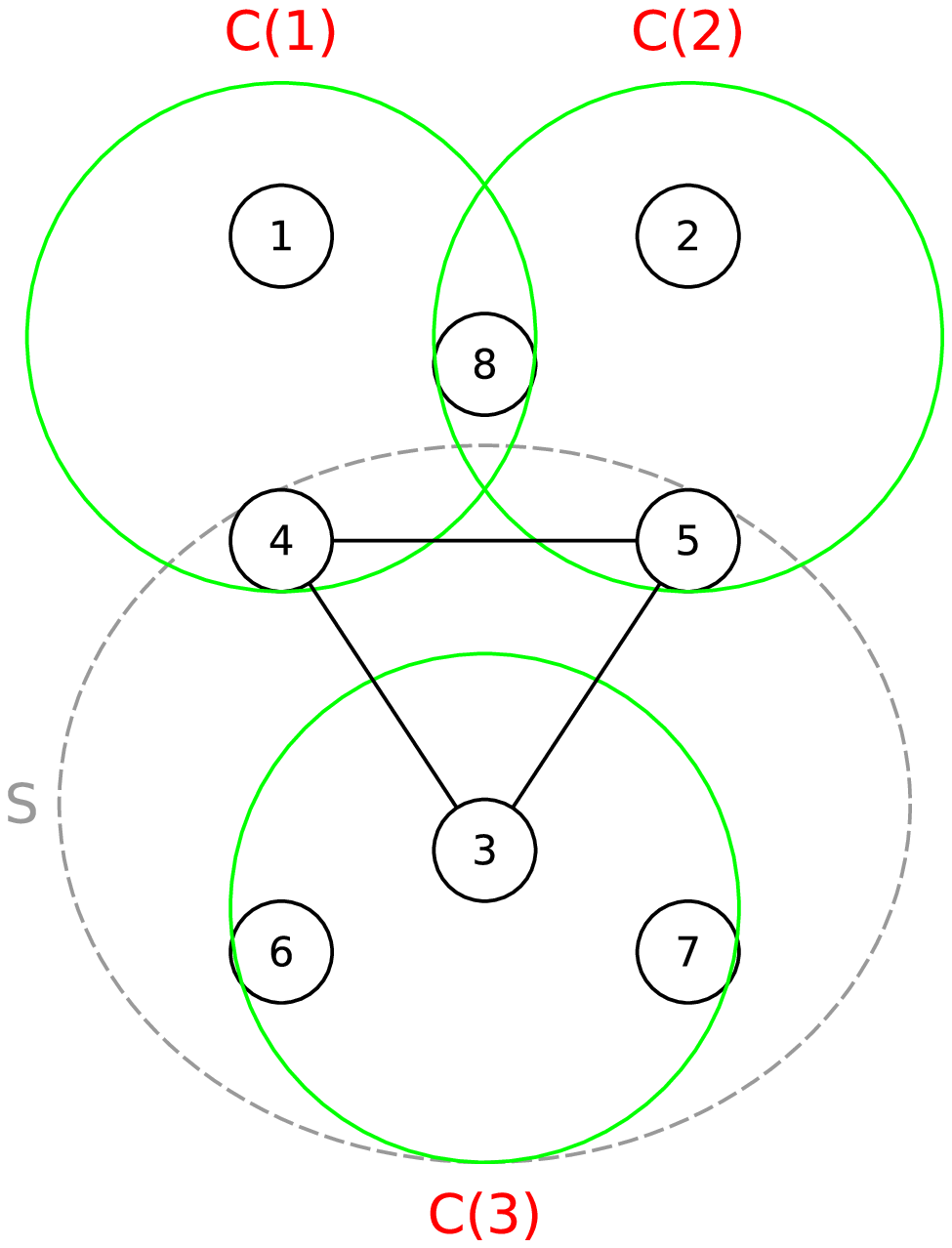}
		\label{fig:validGammaV}}
	\caption{Example of $S \notin \gamma(V)$ (a) and $S \in \gamma(V)$ (b).}
	\label{fig:gammaV}
\end{figure*}

Let $D(S)$ be the union of sets $D(v)$ for all $v \in S$, i.e., $D(S) = \displaystyle \bigcup_{v \in S}{D(v)}$ and let $\gamma(V)$ be the family of all the subsets of vertices that contains $C(v)$ for at least one vertex $v \in V$, i.e., $\gamma(V) = \{F \subseteq \mathcal{P}(V): \forall S \in F, \exists v \in S, C(v) \subseteq S\}$ where $\mathcal{P}(V)$ is the power set of $V$. To exemplify the concept of $\gamma(V)$, consider sets $C(1) = \{1, 4, 8\}$, $C(2) = \{2, 5, 8\}$ and $C(3) = \{3, 6, 7\}$ as shown in Figure~\ref{fig:gammaV}. As exemplified in Figure~\ref{fig:invalidGammaV}, if $S = \{3, 4, 5\}$, then none of the sets $C(1)$, $C(2)$ and $C(3)$ is a subset of $S$, thus $S \notin \gamma(V)$. However, if $S = \{3, 4, 5, 6, 7\}$, then set $C(3)$ is contained in $S$, thus $S \in \gamma(V)$, as shown in Figure~\ref{fig:validGammaV}. The following family of inequalities are valid for the CSP:

\begin{align}
& \sum_{e \in \delta(S)}{x_e} \geqslant 2 \qquad & \forall S \in \gamma(V): D(S) \neq V, \label{eq:inequalityOneCSP} \\
& \sum_{e \in \delta(S)}{x_e} \geqslant 2 y_i \qquad & \forall S \not\in \gamma(V): D(S) \neq V, i \in S,  \label{eq:inequalityTwoCSP} \\
& \sum_{e \in \delta(S)}{x_e} \geqslant 2 (y_i + y_j - 1) \qquad & \forall S \not\in \gamma(V): D(S) = V, i \in S, j \in V \backslash S. \label{eq:inequalityThreeCSP}
\end{align}

Inequalities (\ref{eq:inequalityOneCSP}) ensure that each cut separating two sets $C(v)$ and $C(w)$ must be crossed at least twice. Inequalities (\ref{eq:inequalityTwoCSP}) imply that each cut separating one vertice in the tour and one set $C(v)$ must be crossed at least twice. Inequalities (\ref{eq:inequalityThreeCSP}) ensure that each cut separating two vertices in the tour must be crossed at least twice.
Originally in GTSP, inequalities \eqref{eq:inequalityOneCSP}, \eqref{eq:inequalityTwoCSP}, and \eqref{eq:inequalityThreeCSP} were applied to every subset of vertices containing at least one cluster, i.e., any subset of family $\gamma(V)$.

In the following, a new family of valid inequalities is proposed to consider a scenario particular to the CSP.

\subsection{Cover Intersection Inequalities}

Consider the case in which two covering sets $C(v)$ and $C(u)$, for some pair of vertices $v$ and $u$, overlap. This is a typical scenario for the CSP, and it does not occur on the GTSP since in that problem the clusters are disjoint. The new valid inequalities extend the idea of inequalities \eqref{eq:inequalityOneCSP}, in the sense of requiring a minimum weight for any edge cut-set separating two covering sets. However, to address the overlap of covering sets, the new valid inequalities \eqref{eq:newInequalityCSP} also take into account the edge cut-set weight of the intersection $C(v) \cap C(u)$.


\begin{figure*}[htbp]
  \centering
  \subfloat[Feasible solution]{	
		\includegraphics[width=0.35\textwidth]{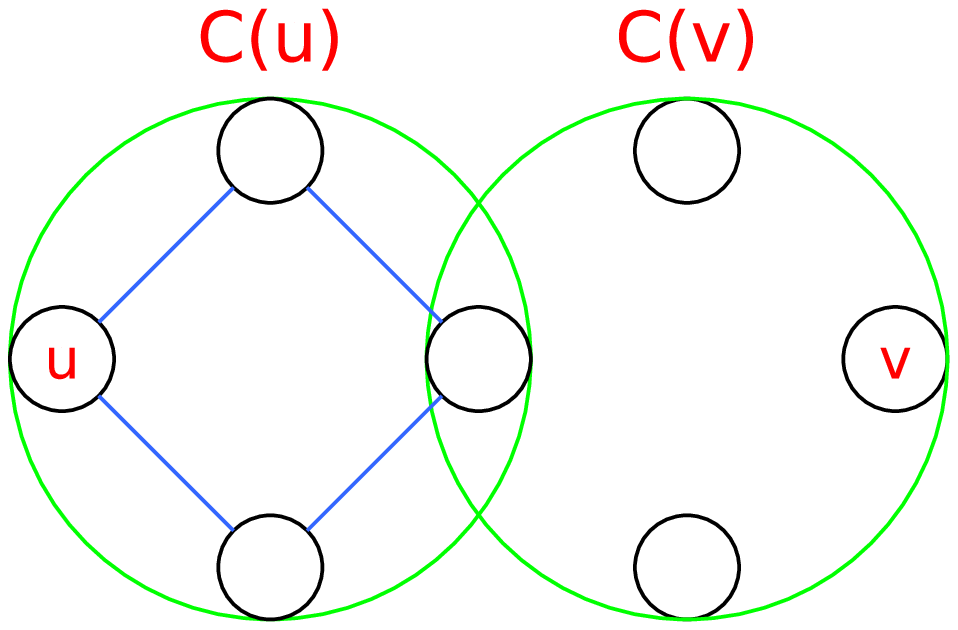}
		\label{fig:coveringOverlapA} }
		\hspace{0.25cm}
  \subfloat[Infeasible solution]{	
		\includegraphics[width=0.35\textwidth]{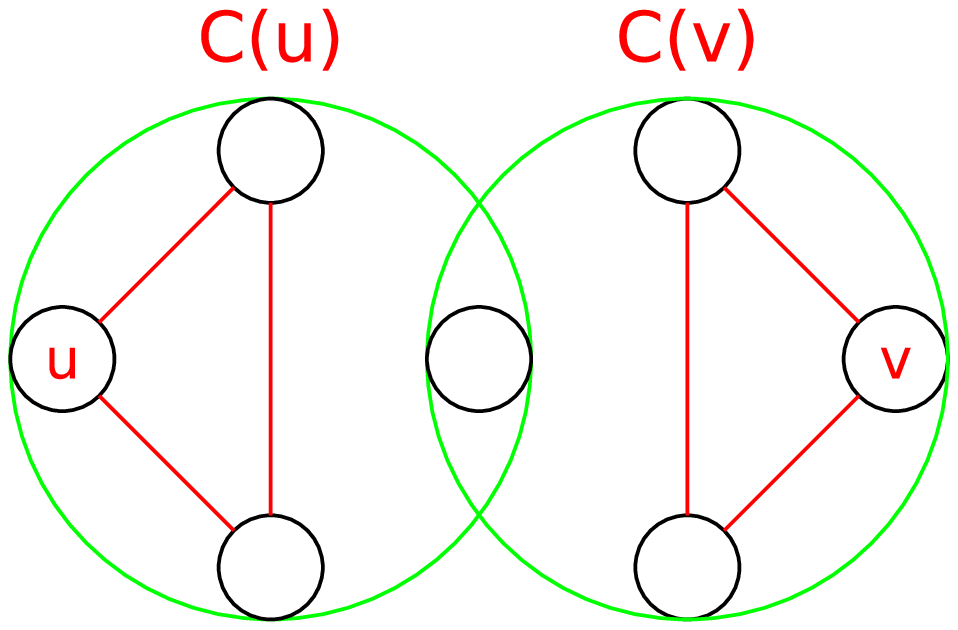}
		\label{fig:coveringOverlapB}}
	\caption{Example of feasible (a) and infeasible (b) solutions in the context of overlap of covering sets.}
	\label{fig:coveringOverlap}
\end{figure*}

For the following new valid inequalities \eqref{eq:newInequalityCSP}, consider $S_{v} = S \cap C(v)$ for any $v \in V$. These inequalities are here called \textit{CI inequalities} (cover intersection inequalities), and they only require a proper subset $S \subset V$ such that $S \in \gamma(V)$, which means it can be employed even if $D(S) = V$, another case in which inequalities \eqref{eq:inequalityOneCSP} cannot be employed.
\begin{align}
& (CI \ inequalities) \nonumber \\
& \sum_{e \in ( \delta(S) \bigcup \delta(S_v))}{x_e} \geqslant 2 \qquad & \forall v \in V, \forall S \subset V : S \in \gamma(V) \label{eq:newInequalityCSP}
\end{align}

According to constraints $(3)$, for any given vertex $v$, at least one vertex of $C(v)$ must be visited by the tour. In other words, for any subset $S \subset V$ such that $S \in \gamma(V)$, the tour must visit $S_v$ or $C(v) \setminus S_v$. If set $S$ does not intersect with $C(v)$, then $S_{v}$ is empty, and \eqref{eq:newInequalityCSP} reduces to \eqref{eq:inequalityOneCSP}. Otherwise, $S_{v}$ is not empty, and in this case, to satisfy constraints $(3)$, the solution must contain at least two edges in either $\delta(S_v)$ or $\delta(S \setminus S_v)$.
In Figure~\ref{fig:coveringOverlapA}, a feasible solution is presented in which both covering sets $C(u)$ and $C(v)$ are visited by the same tour. Despite the fact that the edge cut-set $\delta(C(u))$ is empty is not a concern, since $\delta(C(u) \cap C(v))$ contains two edges. This is not the case in Figure~\ref{fig:coveringOverlapB}, where both $\delta(C(u))$ and $\delta(C(u) \cap C(v))$ are empty, asserting the infeasibility. It is worth mentioning that the solution depicted in  Figure~\ref{fig:coveringOverlapB} violates the CI inequality associated to the set $S = C(u)$ and vertex $v$. 

\section{Branch-and-cut framework} \label{sec:cutSeparation}

This section presents the separation routines for inequalities \eqref{eq:inequalityOneCSP}-\eqref{eq:newInequalityCSP}. Sections \ref{ssec:integer_sep} and \ref{ssec:frac_sep} present the separation routines for integer and fractional solutions, respectively. In the following sections, consider $\mathbf{\{x^I,y^I\}}$ and $\mathbf{\{x^F,y^F\}}$ as integer and fractional solutions for the CSP formulation without the subcycle elimination constraints  \eqref{eq:sec} but possibly including some of the valid inequalities \eqref{eq:inequalityOneCSP}-\eqref{eq:newInequalityCSP}. Also, let $G^I(V^I,E^I)$ and $G^F(V^F,E^F)$ be the graphs induced by $\mathbf{\{x^I,y^I\}}$ and $\mathbf{\{x^F,y^F\}}$, respectively. In $G^I$, every vertex $v \in V^I$ has a weight $y_v$, such that $y_v \in \mathbf{y^I}$, and every edge $e \in E^I$ has a cost $x_e$, such that $x_e \in \mathbf{x^I}$. Similarly, in $G^F$, every vertex $v \in V^F$ has a weight $y_v$, such that $y_v \in \mathbf{y^F}$, and every edge $e \in E^F$ has a cost $x_e$, such that $x_e \in \mathbf{x^F}$.

\subsection{Separation routine for integer solutions} \label{ssec:integer_sep}

The proposed separation routine searches, in a lazy constraint fashion, for inequalities (\ref{eq:inequalityOneCSP}-\ref{eq:newInequalityCSP}) that are possibly violated by an integer solution $\mathbf{\{x^I,y^I\}}$. First, the routine performs a depth-first search in $G^I$ to check for the existence of illegal subcycles.

Let $S \subset V$ be the vertices of an illegal subcycle in $G^I$. To apply inequality \eqref{eq:inequalityOneCSP} or \eqref{eq:newInequalityCSP} with respect to set $S$, it is necessary that $S \in \gamma(V)$. If this is not the case, the proposed routine attempts to augment $S$ into $S_{aug}$ by including the set $C(v)$ for some $v \in S$. However, the choice of which $C(v)$ will be included in $S_{aug}$ is relevant to the effectiveness of the corresponding inequalities, as will be explained next.

\begin{figure}[ht]
	\centering
	\includegraphics[width=0.65\textwidth]{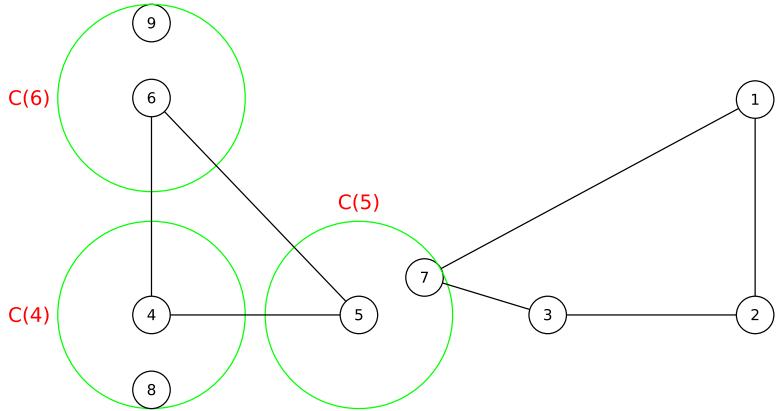} \\ 
	\caption{Example of an invalid CSP solution with two subcycles.}
	\label{fig:vizinhancaCompletaValida}
\end{figure}

Consider Figure~\ref{fig:vizinhancaCompletaValida}, which shows a solution formed by two subcycles in graph $G^I$. In this figure $C(v_4) = \{v_4,v_8\}$, $C(5) = \{v_5, v_7\}$, and $C(v_6) = \{v_6,v_9\}$. By taking the illegal subcycle represented by $S = \{v_4, v_5, v_6\}$, it is not possible to apply inequalities \eqref{eq:inequalityOneCSP} or \eqref{eq:newInequalityCSP}, since $S \notin \gamma(V)$. By taking $S_{aug} = S \cup C(5)$, then $S_{aug} \in \gamma(V)$, however $S_{aug}$ would not generate an effective cut, since vertex $v_7 \in V^I$. Otherwise, effective cuts can be derived from $S_{aug} = S \cup C(v_4)$ or $S_{aug} = S \cup C(v_6)$. Therefore, for an inequality \eqref{eq:inequalityOneCSP} or \eqref{eq:newInequalityCSP} associated to $S_{aug}$ to be effective in cutting solution $\mathbf{\{x^I,y^I\}}$, set $S_{aug}$ cannot contain any vertex in $V^I \setminus S$. 

Algorithm~\ref{alg:separacaoInteira} presents the implementation details of the separation routine for integer solutions, which searches for inequalities (\ref{eq:inequalityOneCSP}-\ref{eq:newInequalityCSP}) associated to each subcycle found in $\mathbf{\{x^I,y^I\}}$. The overall complexity of Algorithm~\ref{alg:separacaoInteira} is bounded by $O(V^2)$.

\begin{algorithm}
\hspace*{\algorithmicindent} \textbf{Input:} graph $G^I(V^I,E^I)$ induced by an infeasible integer solution $\mathbf{\{x^I,y^I\}}$ for the CSP. \\
\hspace*{\algorithmicindent} \textbf{Output:} a set $T$ of valid inequalities that cuts $\mathbf{\{x^I,y^I\}}$.
\begin{algorithmic}[1]
    \For{each subcycle $S$ in $\mathbf{\{x^I,y^I\}}$}
        \State $T \gets \emptyset$
        \If{$D(S) \neq V$}
            \If{$S \in \gamma(V)$}
                \State  $T \gets T \cup $ inequality (\ref{eq:inequalityOneCSP}) associated to $S$
            \Else
                \State  $T \gets T \cup $ inequality (\ref{eq:inequalityTwoCSP}) associated to $S$
                \For{each $v \in S$}
	        		\State $S_{aug} \gets S \cup C(v)$
	            	\If{$S_{aug} \cap (V^I \setminus S) = \emptyset$}
                        \If{$D(S_{aug}) \neq V$}
                            \State  $T \gets T \cup $ inequality (\ref{eq:inequalityOneCSP}) associated to $S_{aug}$
                        \Else
                            \For{$u \in V$}
                                \State $S_{u} \gets S \cap C(u)$;
                                \If{$\delta(S_{u}) \cap E^I = \emptyset$}
                                    \State  $T \gets T \cup $ CI inequality (\ref{eq:newInequalityCSP}) associated to $S_{aug}$ and $S_{u}$
                                \EndIf
                            \EndFor
                        \EndIf
	            	\EndIf
	        	\EndFor
            \EndIf
        \Else
            \For{each subcycle $S'$ in $\mathbf{\{x^I,y^I\}}: S' \neq S$}
                \State  $T \gets T \cup $ inequality (\ref{eq:inequalityThreeCSP}) associated to $S$ and $S'$
            \EndFor
        \EndIf
    \EndFor
    \State \Return $T$;
\end{algorithmic}
\caption{Separation routine for integer solutions.}
\label{alg:separacaoInteira}
\end{algorithm}

\subsection{Exact separation routine for fractional solutions} \label{ssec:frac_sep}




This section gives the exact separation routines of inequalities (\ref{eq:inequalityOneCSP}-\ref{eq:newInequalityCSP}) for a fractional CSP solution $\mathbf{\{x^F,y^F\}}$. In particular, the separation of inequalities (\ref{eq:inequalityOneCSP}), (\ref{eq:inequalityTwoCSP}) and (\ref{eq:inequalityThreeCSP}) follows the methodology proposed by Fischetti, González and Toth~\cite{fischetti1997branch} for the GTSP. As for the CI inequalities (\ref{eq:newInequalityCSP}), a transformation of the solution graph $G'$ is proposed to tackle the overlap of covering sets. The routines are described next.

\begin{figure}[!]
	\centering
	\includegraphics[width=0.65\textwidth]{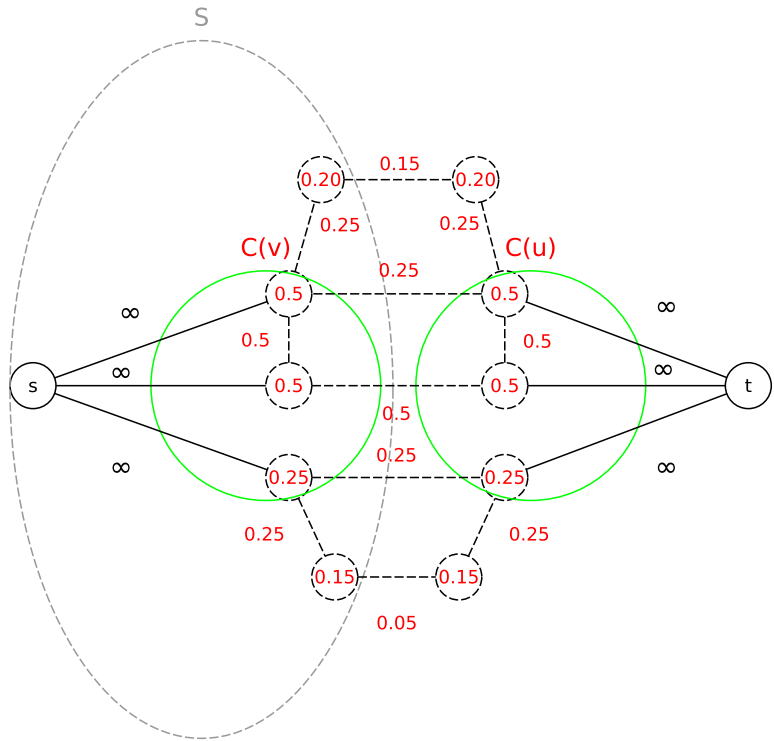}
	\vspace*{-0.05cm}
	\caption{Max-flow instance for the separation of inequality (\ref{eq:inequalityOneCSP}) in the case where $C(v) \cap C(u) = \emptyset$.}
    \label{fig:maxFlow}
\end{figure}

\begin{figure*}[htbp]
  \centering
  \subfloat[Pre-augmentation]{	
		\includegraphics[width=0.45\textwidth]{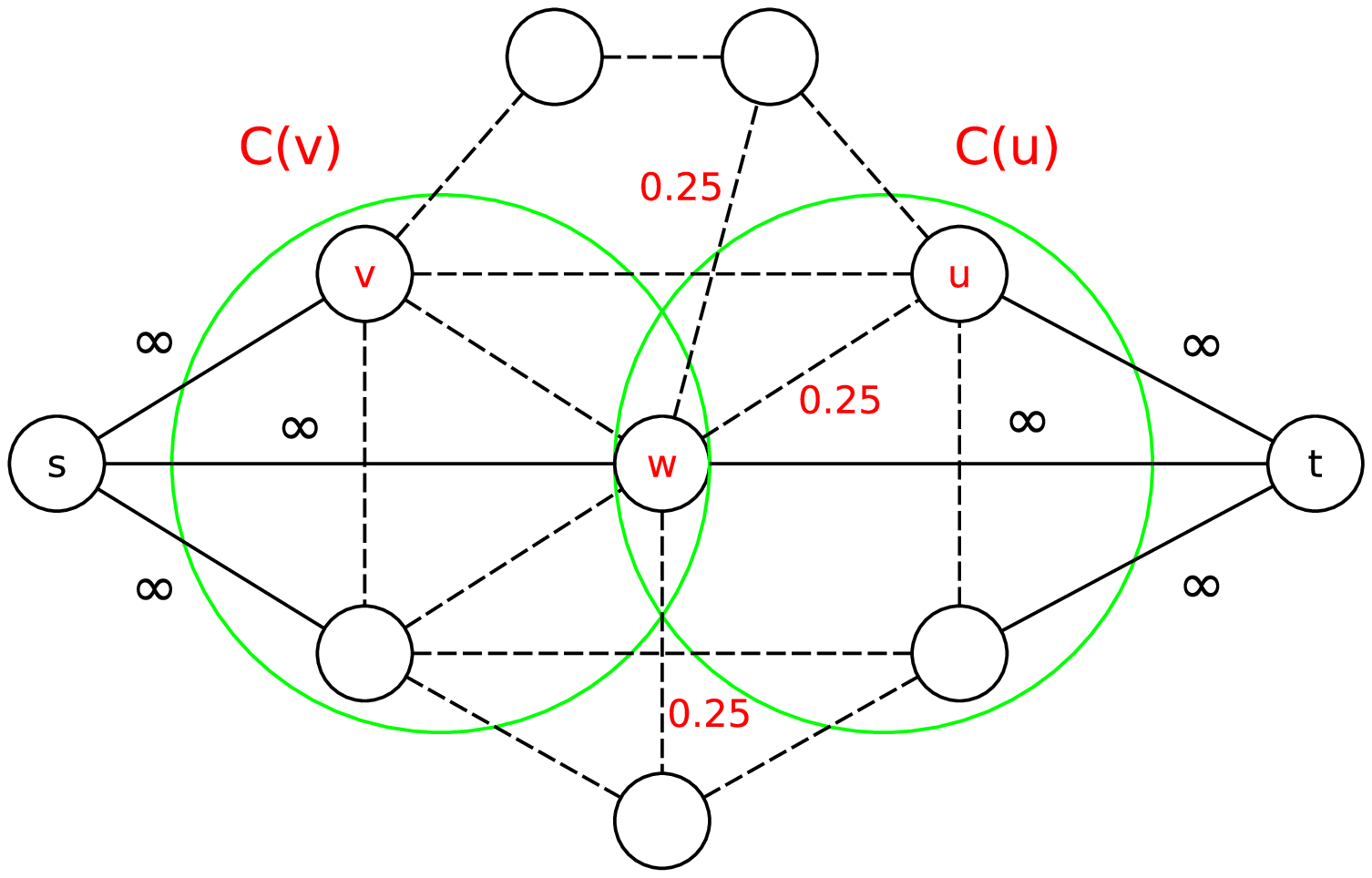}
		\label{fig:graphTransformationA} }
		\hspace{0.5cm}
  \subfloat[Augmented graph]{	
		\includegraphics[width=0.45\textwidth]{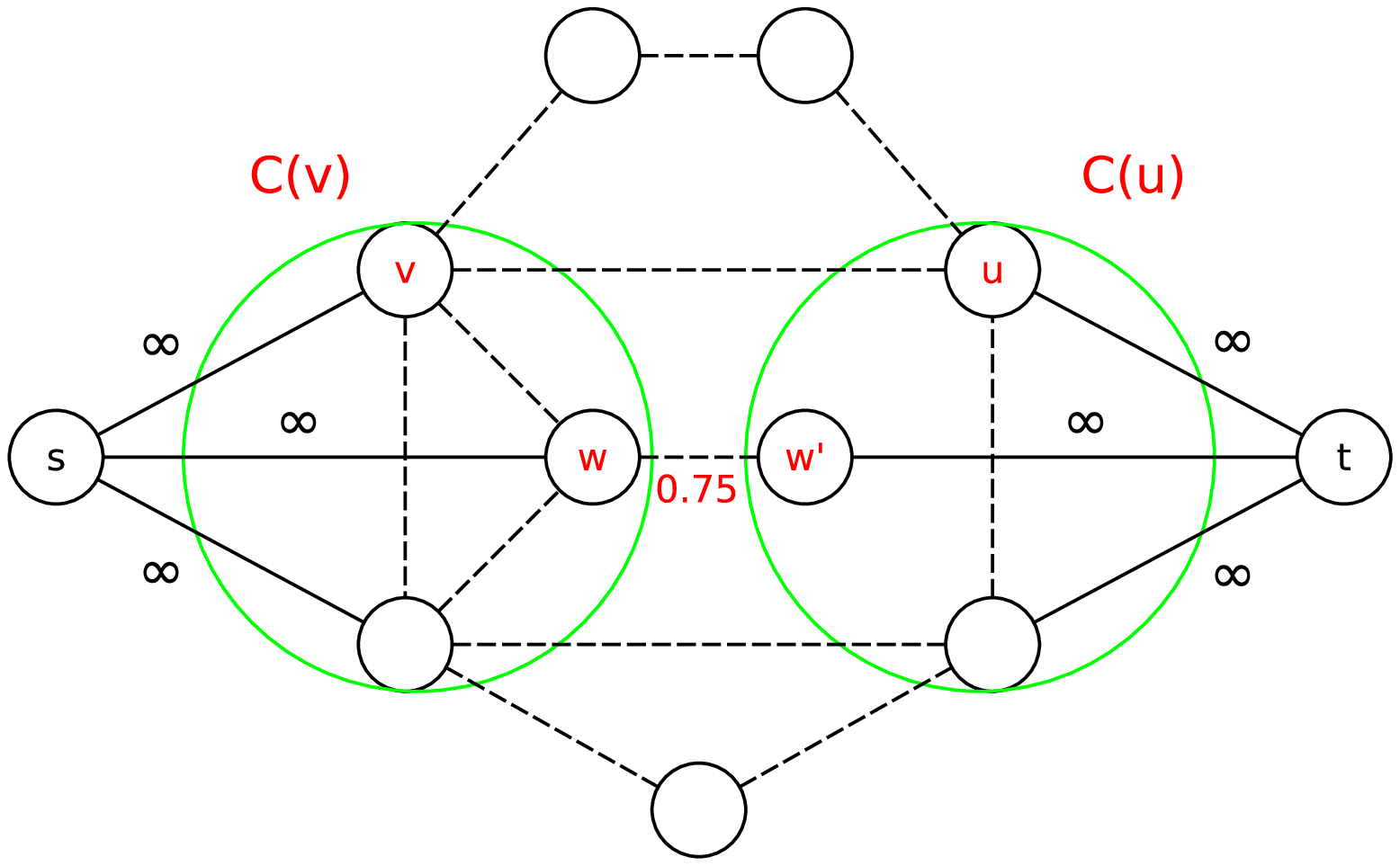}
		\label{fig:graphTransformationB}}
	\caption{Graph augmentation for the separation of CI inequalities~\eqref{eq:newInequalityCSP}.}
	\label{fig:graphTransformation}
\end{figure*}

As observed by Fischetti, González and Toth~\cite{fischetti1997branch}, the separation problem to find one or more inequalities (\ref{eq:inequalityThreeCSP}) violated by  $\mathbf{\{x^F,y^F\}}$ can be reduced to the problem of computing a minimum cut between two vertices $i$ and $j$ in graph $G^F$, $i \in S$ and $j \in V^F \backslash S$, i.e., finding the maximum flow from $i$ to $j$ \cite{ahuja1993network}.
Similarly, the separation of inequalities (\ref{eq:inequalityTwoCSP}) can be reduced to computing a minimum cut in graph $G^F$ that separates $i \in S$ and $C(u) \subseteq V^F \backslash S$. In other words, finding the maximum flow from $i$ to $t$ \cite{ahuja1993network}, where $t$ is an artificial vertex connected to each $j \in C(u)$ through edges with infinite capacity.
As for inequalities (\ref{eq:inequalityOneCSP}), the separation problem can be reduced to computing a minimum cut between covering sets $C(v)$ and $C(u)$ in graph $G^F$, with $C(v) \subseteq S$, $C(u) \subseteq V \backslash S$, and $C(v) \cap C(u) = \emptyset$. A maximum flow from $s$ to $t$ can be computed, such that $s$ and $t$ are artificial vertices connected, respectively, to each vertex in $C(v)$ and $C(u)$ with infinite capacity edges, as illustrated in Figure \ref{fig:maxFlow}. It is worth noting that the separation of inequality (\ref{eq:inequalityOneCSP}) does not work when $C(v)$ and $C(u)$ overlap, since every cut separating $s$ and $t$ has infinite weight, as exemplified in Figure \ref{fig:graphTransformationA}. 

An exact separation algorithm for CI inequalities \eqref{eq:newInequalityCSP} is proposed to accomodate the case when two covering sets $C(v)$ and $C(u)$ overlap.
The first step is to augment graph $G^F$, by including an artificial vertex $w'$ and an artificial edge $(w,w')$ for every vertex $w \in C(v) \cap C(u)$. Vertex $w$ is removed from $C(u)$ and vertex $w'$ is included into $C(u)$. Finally, for each $w \in C(v) \cap C(u)$, let $T_w$ be the set of edges with one endpoint being $w$ and the other is in $V \backslash C(v)$. The edges of $T_w$ are excluded from $G^F$ and their total weight is transferred to the artificial edge $(w,w')$. This ensures that every artificial edge will be counted for in any minimum cut, in the sense that every edge in $T_w$ contributes in their purpose of connecting both covering sets $C(v)$ and $C(u)$, as expected in a feasible solution. Figure~\ref{fig:graphTransformation} illustrates the augmentation of graph $G^F$. 


The separation of a CI inequality (\ref{eq:newInequalityCSP}) reduces to computing a minimum cut between sets $C(v)$ and $C(u)$ in the augmented graph. 
Let $\delta(S_{min})$ be the minimum cut between $C(v)$ and $C(u)$ and $S_{u} = S_{min} \cap C(u)$. If $\displaystyle \sum_{e \in \delta(S_{min}) \bigcup \delta(S_u)}{x_e}$ has a value less than $2$, then a violated CI inequality (\ref{eq:newInequalityCSP}) was found.

Algorithm~\ref{alg:separacaoExataFullFracionaria} presents the implementation of exact separation routine for fractional solutions. The separation consists in computing a max-flow for each pair of vertices, thus considering a push-relabel algorithm \cite{ahuja1993network} to solve max-flow, the time complexity of Algorithm~\ref{alg:separacaoExataFullFracionaria} is bounded by $O(V^4E)$.

\begin{algorithm}
\hspace*{\algorithmicindent} \textbf{Input:} graph $G^F(V^F,E^F)$ induced by a fractional solution $\mathbf{\{x^F,y^F\}}$ for the CSP. \\
\hspace*{\algorithmicindent} \textbf{Output:} a set $T$ of valid inequalities that cuts $\mathbf{\{x^F,y^F\}}$.
\begin{algorithmic}[1]
    \State $T \gets \emptyset$
    \For{$v \in V$}
        \For{$u \in V \setminus \{v\}$}
            \If{$D(C(v)) \neq V$ \textbf{and} $C(v) \cap C(u) = \emptyset$}
                \State $S \gets$ minCut($C(v),C(u),G^F$)
                \State $T \gets T \cup$ inequality (\ref{eq:inequalityOneCSP}) associated to $S$.
            \Else
                \State $G_{aug}^F \gets$ augment($G^F$)
                \State $S \gets$ minCut($C(v),C(u),G_{aug}^F$)
                \State $T \gets T \cup$ CI inequality (\ref{eq:newInequalityCSP}) associated to $S$ and $u$.
            \EndIf
            \If{$y_v > 0$ \textbf{and} $v \notin C(u)$}
                \State $S \gets$ minCut($v,C(u),G^F$)
                \State $T \gets T \cup$ inequality (\ref{eq:inequalityTwoCSP}) associated to $S$ and $u$.
            \EndIf
            \If{$y_v + y_u - 1 > 0$}
                \State $S \gets$ minCut($v,u,G^F$)
                \State $T \gets T \cup$ inequality (\ref{eq:inequalityThreeCSP}) associated to $S$, $v$ and $u$.
            \EndIf
        \EndFor
    \EndFor
\State \Return $T$;
\end{algorithmic}
\caption{Exact separation routine for fractional solutions.}
\label{alg:separacaoExataFullFracionaria}
\end{algorithm}

Given the computational effort required for the exact separation of fractional solutions, two alternatives were investigated. The first is based on a \textit{first-found} policy, which follows the same steps of Algorithm~\ref{alg:separacaoExataFullFracionaria}, however the execution is interrupted once the first inequality which surpasses a given violation threshold $\epsilon$ is found. For example, with respect to inequalities \eqref{eq:inequalityOneCSP}, given a vertex $v \in V$ and a set $S \in \gamma(V) : D(S) \neq V$, if the following holds, $(2 - \sum_{e \in \delta(S)}{x_e} > \epsilon)$, then the cut is included in the model and Algorithm~\ref{alg:separacaoExataFullFracionaria} halts. The same goes for inequalities~(\ref{eq:inequalityTwoCSP}-\ref{eq:newInequalityCSP}).

The second alternative for the exact separation routines resides in the heuristic separation of inequalities (\ref{eq:inequalityOneCSP}-\ref{eq:newInequalityCSP}), described in the following section.

\subsection{Heuristic separation routine for fractional solutions}

A heuristic separation has the purpose of finding inequalities being violated by a fractional solution $\{\mathbf{x^F,y^F}\}$ within short computational times. In contrast with the exact separation routine however, a heuristic does not come with any guarantee of finding a violated inequality even if one exists.

The heuristic separation routine for inequalities~(\ref{eq:inequalityOneCSP}-\ref{eq:newInequalityCSP}) is composed of four main steps. The first step searches for inequalities (\ref{eq:inequalityOneCSP}) and (\ref{eq:newInequalityCSP}) for every $u \in V$ and its corresponding covering set $C(u)$. In more details, let $S = C(u)$ and consider two cases: $(i)$ if $D(S) \neq V$ and $\sum_{e \in \delta(S)}{x_e} < 2$, then the inequality (\ref{eq:inequalityOneCSP}) associated to $S$ cuts $\{\mathbf{x^F,y^F}\}$; $(ii)$ if $D(S) = V$ and $\sum_{e \in \delta(S)}{x_e} + \sum_{e \in \delta(S_{v}) \setminus \delta(S)}{x_e} < 2$ for some vertex $v \in V \setminus \{u\}$, then the CI inequality (\ref{eq:newInequalityCSP}) associated to $S$ and $v$ cuts $\{\mathbf{x^F,y^F}\}$. 

In the second step, the connected components $S_1,\ldots, S_p$ of $G^F$ are computed. For each component $S_k$, let $S = S_k$ and if $S \in \gamma(V)$ then two cases are considered: $(i)$ if $D(S) \neq V$, then the inequality (\ref{eq:inequalityOneCSP}) associated to $S$ cuts  $\{\mathbf{x^F,y^F}\}$; $(ii)$ if $D(S) = V$ and $\sum_{e \in \delta(S_{v})}{x_e} < 2$ for some vertex $v \in V$, then the CI inequality (\ref{eq:newInequalityCSP}) associated to $S$ and $v$ cuts $\{\mathbf{x^F,y^F}\}$.
.

In the third step, for each connected component $S_k$, let $S = S_k$ and $i = \arg\max_{v}\{y_v: v \in S\}$. If $D(S) \neq V$, then the inequality (\ref{eq:inequalityTwoCSP}) associated to $S$ and $v$ cuts $\{\mathbf{x^F,y^F}\}$.

Finally, the fourth step iterates through all pairs of connected components $S_k$ and $S_l$, $k \neq l$. For each pair, let $i = \arg\max_{v}\{y_v: v \in S_k\}$ and $j = \arg\max_{v}\{y_v: v \in S_l\}$. If $y_i + y_j > 1$, the inequality (\ref{eq:inequalityThreeCSP}) associated to $S = S_k$, $i$, and $j$ cuts $\{\mathbf{x^F,y^F}\}$.

Algorithm~\ref{alg:separacaoHeuristicaFracionaria}, with a time complexity bounded by $O(V^2)$, details the heuristic separation routine for fractional solutions.

\begin{algorithm}
\small
\hspace*{\algorithmicindent} \textbf{Input:} graph $G^F(V^F,E^F)$ induced by a fractional solution $\mathbf{\{x^F,y^F\}}$ for the CSP. \\
\hspace*{\algorithmicindent} \textbf{Output:} a set $T$ of valid inequalities that cuts $\mathbf{\{x^F,y^F\}}$.
\begin{algorithmic}[1]
    \State $T \gets \emptyset$
    \For{$u \in V$}
        \State $S \gets C(u)$;
        \If{$D(S) \neq V$}
            \If{\begin{footnotesize}$ \sum_{e \in \delta(S)}{x_e} < 2$\end{footnotesize}}
                \State $T \gets T \cup$ inequality (\ref{eq:inequalityOneCSP}) associated to $S$.
            \EndIf
        \Else
            \For{$v \in V: v \neq u$}
                \State $S_{v} \gets S \cap C(v)$;
                \If{\begin{footnotesize}$\sum_{e \in \delta(S)}{x_e} + \sum_{e \in \delta(S_{v}) \setminus \delta(S)}{x_e} < 2$\end{footnotesize}}
                    \State $T \gets T \cup$ CI inequality (\ref{eq:newInequalityCSP}) associated to $S$ and $v$.
                \EndIf
            \EndFor
        \EndIf
    \EndFor
    \State Compute the connected components $S_1,\ldots,S_p$ of $G^F$;
    \For{$k = 1, ..., p$}
        \State $S \gets S_k$
        \If{$S \in \gamma(V)$}
            \If{$D(S) \neq V$}
                \State $T \gets T \cup$ inequality (\ref{eq:inequalityOneCSP}) associated to $S$.
            \Else
                \For{$v \in V : v \neq w$}
                    \State $S_v \gets S \cap C(v)$;
                    \If{\begin{footnotesize}$\sum_{e \in \delta(S_{v}) \setminus \delta(S)}{x_e} < 2$\end{footnotesize}}
                        \State $T \gets T \cup$ CI inequality (\ref{eq:newInequalityCSP}) associated to $S$ and $v$.
                    \EndIf
                \EndFor
            \EndIf
        \Else
        \State $i \gets \arg\max_{v}\{y_v: v \in S\}$.
        \If{$D(S) \neq V$}
            \State $T \gets T \cup$ inequality (\ref{eq:inequalityTwoCSP}) associated to $S$ and $i$.
        \EndIf
        \For{$l = k, ..., p$}
            \State $j \gets \arg\max_{v}\{y_v: v \in S_l\}$
            \State $T \gets T \cup$ inequality (\ref{eq:inequalityThreeCSP}) associated to $S$, $i$ and $j$.
        \EndFor
        \EndIf
    \EndFor
\State \Return T
\end{algorithmic}
\caption{Heuristic separation routine for fractional solutions.}
\label{alg:separacaoHeuristicaFracionaria}
\end{algorithm}

\section{Computational Experiments} \label{sec:experiments}

In this section, the proposed branch-and-cut methodologies are evaluated and compared to the state-of-the-art using the literature benchmark of instances, described in the following section.

\subsection{Instances}

The set of instances used in the computational experiments was created by Salari et al.\cite{salari2015combining} based on the TSPLIB \cite{reinelt1991tsplib}, and they are divided in two types: small ($36$ instances) and medium ($12$ instances). The small instances have $51 \leqslant |V| \leqslant 100$ vertices and medium-size instances have $150 \leqslant |V| \leqslant 200$ vertices. There is also a set of large instances with $|V| \geqslant 532$, which Salari et al. assigned only for the testing of heuristics, and for this reason, the large instances are not reported in this paper.  The covering set of each vertex is defined by its $k$ closest vertices. For each graph, three values of $k$ were used, $k=7$, $k=9$, and $k=11$. Full experimental data (including results for large instances), instances, and source codes are available on-line\footnote{\label{note1}\url{http://www.ic.unicamp.br/~fusberti/problems/csp}}.

\subsection{Computational Settings}

The branch-and-cut methodologies were implemented in C++ using solver Gurobi, with a one-hour time limit. The experiments were conducted on a PC under Ubuntu and CPU Intel Xeon E5-2630  $2.2$ GHz, with $64$GB of RAM.

\subsection{Evaluated Methodologies}

Five branch-and-cut methodologies were implemented and evaluated in the computational experiments:

\begin{enumerate}
    \item $CSP$-$I$: exact separation routine for integer solutions (Algorithm~\ref{alg:separacaoInteira}) considering valid inequalities (\ref{eq:inequalityOneCSP}), (\ref{eq:inequalityTwoCSP}), and (\ref{eq:inequalityThreeCSP}), but excluding the CI inequalities  \eqref{eq:newInequalityCSP};
    
    \item $CSP$-$I\&F_{vp}$: on the root node, exact separation routine for fractional solutions (Algorithm~\ref{alg:separacaoExataFullFracionaria}) considering  inequalities (\ref{eq:inequalityOneCSP}), (\ref{eq:inequalityTwoCSP}), and (\ref{eq:inequalityThreeCSP}),  but excluding the CI inequalities for CSP \eqref{eq:newInequalityCSP}. For the non-root nodes, Algorithm~\ref{alg:separacaoExataFullFracionaria} was implemented under the \textit{first-found} policy with violation threshold $\epsilon = 1$ (see Section
   ~\ref{ssec:frac_sep}).
   
    \item $CSP$-$I\&F_{vp}$-$X$: same as $CSP$-$I\&F_{vp}$, but including the CI inequalities  (\ref{eq:newInequalityCSP});
    
    \item $CSP$-$I\&F_{h}$: on the root node, exact separation routine for fractional solutions (Algorithm~\ref{alg:separacaoExataFullFracionaria}) considering  inequalities (\ref{eq:inequalityOneCSP}), (\ref{eq:inequalityTwoCSP}), and (\ref{eq:inequalityThreeCSP}),  but excluding the CI inequalities \eqref{eq:newInequalityCSP}. For the non-root nodes, heuristic separation for fractional solutions (Algorithm
   ~\ref{alg:separacaoHeuristicaFracionaria}, considering  inequalities (\ref{eq:inequalityOneCSP}), (\ref{eq:inequalityTwoCSP}), and (\ref{eq:inequalityThreeCSP}), but excluding the CI inequalities  \eqref{eq:newInequalityCSP};
    
    \item $CSP$-$I\&F_{h}$-$X$: Same as $CSP$-$I\&F_{h}$, but including inequalities  (\ref{eq:newInequalityCSP});
\end{enumerate}



These methodologies were compared with the integer linear programming formulation proposed by Salari et al.\cite{salari2015combining}, denoted here as $SRS$. To the best of our knowledge, $SRS$ is the best performing exact methodology for the CSP.

Preliminary experiments have shown that even in cases where the heuristic separation fails to find violated inequalities in methodologies $CSP$-$I\&F_{h}$ and $CSP$-$I\&F_{h}$-$X$, applying the exact separation does not improve the quality of the solutions obtained. This can be justified by the high computational effort spent by the exact separation routines.

\subsection{Results}

The results of the computational experiments are reported for the small and medium instances in Tables~\ref{tab:smallInstances} and~\ref{tab:mediumInstances}, respectively. Each table reports for each methodology and for each instance, the following:

\begin{itemize}
	\item \textit{LB}: best lower bound obtained;
	\item \textit{Gap}: optimality gap $(\displaystyle \frac{UB - LB}{UB}) \cdot 100$;
	\item \textit{Time}: execution time in seconds.
\end{itemize}

In both tables, the column group \textit{BestUB} reports the best upper bounds known in the literature for each instance:  column UB gives the best known upper bounds and column \textit{References} cites the papers which attained them.
For each instance, the Tables~\ref{tab:smallInstances} and~\ref{tab:mediumInstances} highlight the optimal solutions (underlined) and the best lower bounds (in bold) obtained by each methodology. 

\newpage
\thispagestyle{empty}
\newgeometry{top=10mm, bottom=10mm} 
\begin{landscape}
\setlength{\tabcolsep}{2pt}
\scriptsize
\begin{longtable}{lllllllllllllllllllllllllllll}
\label{tab:smallInstances} \\
\caption{Results of computacional experiments for the small-size instances.} \\

&           &  & \multicolumn{2}{l}{BestUB}                                                                                         &  & \multicolumn{3}{l}{$SRS$}                              &  & \multicolumn{3}{l}{$CSP$-$I$}                         &  & \multicolumn{3}{l}{$CSP$-$I\&F_{vp}$}             &  & \multicolumn{3}{l}{$CSP$-$I\&F_{h}$}                  &  & \multicolumn{3}{l}{$CSP$-$I\&F_{vp}$-$X$}            &  & \multicolumn{3}{l}{$CSP$-$I\&F_{h}$-$X$}              \\ \cline{1-2} \cline{4-5} \cline{7-9} \cline{11-13} \cline{15-17} \cline{19-21} \cline{23-25} \cline{27-29}
Instance     & NC        &  & UB        & References                                                                                           &  & LB               & Gap          & Time             &  & LB               & Gap          & Time            &  & LB               & Gap          & Time           &  & LB               & Gap          & Time            &  & LB               & Gap          & Time           &  & LB               & Gap          & Time            \\ \cline{1-2} \cline{4-5} \cline{7-9} \cline{11-13} \cline{15-17} \cline{19-21} \cline{23-25} \cline{27-29}
\endfirsthead

\multicolumn{15}{l}%
{{ \textit{Continued from previous page}}} \\ \hline
&           &  & \multicolumn{2}{l}{BestUB}                                                                                         &  & \multicolumn{3}{l}{$SRS$}                              &  & \multicolumn{3}{l}{$CSP$-$I$}                         &  & \multicolumn{3}{l}{$CSP$-$I\&F_{vp}$}             &  & \multicolumn{3}{l}{$CSP$-$I\&F_{h}$}                  &  & \multicolumn{3}{l}{$CSP$-$I\&F_{vp}$-$X$}            &  & \multicolumn{3}{l}{$CSP$-$I\&F_{h}$-$X$}              \\ \cline{1-2} \cline{4-5} \cline{7-9} \cline{11-13} \cline{15-17} \cline{19-21} \cline{23-25} \cline{27-29}
Instance     & NC        &  & UB        & References                                                                                           &  & LB               & Gap          & Time             &  & LB               & Gap          & Time            &  & LB               & Gap          & Time           &  & LB               & Gap          & Time            &  & LB               & Gap          & Time           &  & LB               & Gap          & Time            \\ \cline{1-2} \cline{4-5} \cline{7-9} \cline{11-13} \cline{15-17} \cline{19-21} \cline{23-25} \cline{27-29}
\endhead

\hline \multicolumn{15}{l}{{\textit{Continued on next page}}} \\
\endfoot

\endlastfoot

eil51        & 7         &  & 164       & \cite{golden2012generalized,salari2012integer,salari2015combining,venkatesh2019multi} &  & \textbf{164}     & \underline{\textbf{0}} & 149              &  & \textbf{164}     & \underline{\textbf{0}} & 2               &  & \textbf{164}     & \underline{\textbf{0}} & 4              &  & \textbf{164}     & \underline{\textbf{0}} & 3               &  & \textbf{164}     & \underline{\textbf{0}} & 3              &  & \textbf{164}     & \underline{\textbf{0}} & 3               \\
             & 9         &  & 159       & \cite{golden2012generalized,salari2012integer,salari2015combining,venkatesh2019multi} &  & \textbf{159}     & \underline{\textbf{0}} & 220              &  & \textbf{159}     & \underline{\textbf{0}} & 1               &  & \textbf{159}     & \underline{\textbf{0}} & 2              &  & \textbf{159}     & \underline{\textbf{0}} & 2               &  & \textbf{159}     & \underline{\textbf{0}} & 4              &  & \textbf{159}     & \underline{\textbf{0}} & 3               \\
             & 11        &  & 147       & \cite{golden2012generalized,salari2012integer,salari2015combining,venkatesh2019multi} &  & \textbf{147}     & \underline{\textbf{0}} & 681              &  & \textbf{147}     & \underline{\textbf{0}} & 1               &  & \textbf{147}     & \underline{\textbf{0}} & 2              &  & \textbf{147}     & \underline{\textbf{0}} & 2               &  & \textbf{147}     & \underline{\textbf{0}} & 5              &  & \textbf{147}     & \underline{\textbf{0}} & 4               \\
             &           &  &           &                                                                                                        &  &                  &                  &                  &  &                  &                  &                 &  &                  &                  &                &  &                  &                  &                 &  &                  &                  &                &  &                  &                  &                 \\
berlin52     & 7         &  & 3887      & \cite{golden2012generalized,salari2012integer,salari2015combining,venkatesh2019multi} &  & \textbf{3887}    & \underline{\textbf{0}} & 140              &  & \textbf{3887}    & \underline{\textbf{0}} & 2               &  & \textbf{3887}    & \underline{\textbf{0}} & 2              &  & \textbf{3887}    & \underline{\textbf{0}} & 2               &  & \textbf{3887}    & \underline{\textbf{0}} & 4              &  & \textbf{3887}    & \underline{\textbf{0}} & 3               \\
             & 9         &  & 3430      & \cite{golden2012generalized,salari2012integer,salari2015combining,venkatesh2019multi} &  & \textbf{3430}    & \underline{\textbf{0}} & 212              &  & \textbf{3430}    & \underline{\textbf{0}} & 1               &  & \textbf{3430}    & \underline{\textbf{0}} & 3              &  & \textbf{3430}    & \underline{\textbf{0}} & 2               &  & \textbf{3430}    & \underline{\textbf{0}} & 6              &  & \textbf{3430}    & \underline{\textbf{0}} & 3               \\
             & 11        &  & 3262      & \cite{golden2012generalized,salari2012integer,salari2015combining,venkatesh2019multi} &  & \textbf{3262}    & \underline{\textbf{0}} & 255              &  & \textbf{3262}    & \underline{\textbf{0}} & 1               &  & \textbf{3262}    & \underline{\textbf{0}} & 2              &  & \textbf{3262}    & \underline{\textbf{0}} & 2               &  & \textbf{3262}    & \underline{\textbf{0}} & 4              &  & \textbf{3262}    & \underline{\textbf{0}} & 4               \\
             &           &  &           &                                                                                                        &  &                  &                  &                  &  &                  &                  &                 &  &                  &                  &                &  &                  &                  &                 &  &                  &                  &                &  &                  &                  &                 \\
st70         & 7         &  & 288       & \cite{golden2012generalized,salari2012integer,salari2015combining,venkatesh2019multi} &  & \textbf{288}     & \underline{\textbf{0}} & 490              &  & \textbf{288}     & \underline{\textbf{0}} & 3               &  & \textbf{288}     & \underline{\textbf{0}} & 7              &  & \textbf{288}     & \underline{\textbf{0}} & 5               &  & \textbf{288}     & \underline{\textbf{0}} & 7              &  & \textbf{288}     & \underline{\textbf{0}} & 7               \\
             & 9         &  & 259       & \cite{golden2012generalized,salari2012integer,salari2015combining,venkatesh2019multi} &  & \textbf{259}     & \underline{\textbf{0}} & 1391             &  & \textbf{259}     & \underline{\textbf{0}} & 3               &  & \textbf{259}     & \underline{\textbf{0}} & 7              &  & \textbf{259}     & \underline{\textbf{0}} & 6               &  & \textbf{259}     & \underline{\textbf{0}} & 13             &  & \textbf{259}     & \underline{\textbf{0}} & 9               \\
             & 11        &  & 247       & \cite{golden2012generalized,salari2012integer,salari2015combining,venkatesh2019multi} &  & 218              & 13.14            & 3600             &  & \textbf{247}     & \underline{\textbf{0}} & 3               &  & \textbf{247}     & \underline{\textbf{0}} & 7              &  & \textbf{247}     & \underline{\textbf{0}} & 5               &  & \textbf{247}     & \underline{\textbf{0}} & 14             &  & \textbf{247}     & \underline{\textbf{0}} & 9               \\
             &           &  &           &                                                                                                        &  &                  &                  &                  &  &                  &                  &                 &  &                  &                  &                &  &                  &                  &                 &  &                  &                  &                &  &                  &                  &                 \\
eil76        & 7         &  & 207       & \cite{golden2012generalized,salari2012integer,salari2015combining,venkatesh2019multi} &  & 193              & 7.45             & 3600             &  & \textbf{207}     & \underline{\textbf{0}} & 6               &  & \textbf{207}     & \underline{\textbf{0}} & 29             &  & \textbf{207}     & \underline{\textbf{0}} & 9               &  & \textbf{207}     & \underline{\textbf{0}} & 15             &  & \textbf{207}     & \underline{\textbf{0}} & 12              \\
             & 9         &  & 185       & \cite{salari2012integer}                                                              &  & 161              & 14.65            & 3600             &  & \textbf{185}     & \underline{\textbf{0}} & 10              &  & \textbf{185}     & \underline{\textbf{0}} & 10             &  & \textbf{185}     & \underline{\textbf{0}} & 9               &  & \textbf{185}     & \underline{\textbf{0}} & 13             &  & \textbf{185}     & \underline{\textbf{0}} & 12              \\
             & 11        &  & 170       & \cite{golden2012generalized,salari2012integer,salari2015combining,venkatesh2019multi} &  & 145              & 17.08            & 3600             &  & \textbf{170}     & \underline{\textbf{0}} & 6               &  & \textbf{170}     & \underline{\textbf{0}} & 8              &  & \textbf{170}     & \underline{\textbf{0}} & 7               &  & \textbf{170}     & \underline{\textbf{0}} & 15             &  & \textbf{170}     & \underline{\textbf{0}} & 13              \\
             &           &  &           &                                                                                                        &  &                  &                  &                  &  &                  &                  &                 &  &                  &                  &                &  &                  &                  &                 &  &                  &                  &                &  &                  &                  &                 \\
pr76         & 7         &  & 50275     & \cite{golden2012generalized,salari2012integer,salari2015combining,venkatesh2019multi} &  & \textbf{50275}   & \underline{\textbf{0}} & 2488             &  & \textbf{50275}   & \underline{\textbf{0}} & 7               &  & \textbf{50275}   & \underline{\textbf{0}} & 8              &  & \textbf{50275}   & \underline{\textbf{0}} & 6               &  & \textbf{50275}   & \underline{\textbf{0}} & 12             &  & \textbf{50275}   & \underline{\textbf{0}} & 8               \\
             & 9         &  & 45348     & \cite{golden2012generalized,salari2012integer,salari2015combining,venkatesh2019multi} &  & 42935            & 5.62             & 3600             &  & \textbf{45348}   & \underline{\textbf{0}} & 6               &  & \textbf{45348}   & \underline{\textbf{0}} & 32             &  & \textbf{45348}   & \underline{\textbf{0}} & 10              &  & \textbf{45348}   & \underline{\textbf{0}} & 16             &  & \textbf{45348}   & \underline{\textbf{0}} & 14              \\
             & 11        &  & 43028     & \cite{golden2012generalized,salari2012integer,salari2015combining,venkatesh2019multi} &  & 39022            & 10.27            & 3600             &  & \textbf{43028}   & \underline{\textbf{0}} & 28              &  & \textbf{43028}   & \underline{\textbf{0}} & 17             &  & \textbf{43028}   & \underline{\textbf{0}} & 46              &  & \textbf{43028}   & \underline{\textbf{0}} & 27             &  & \textbf{43028}   & \underline{\textbf{0}} & 28              \\
             &           &  &           &                                                                                                        &  &                  &                  &                  &  &                  &                  &                 &  &                  &                  &                &  &                  &                  &                 &  &                  &                  &                &  &                  &                  &                 \\
rat99        & 7         &  & 486       & \cite{golden2012generalized,salari2012integer,salari2015combining,venkatesh2019multi} &  & 433              & 12.21            & 3600             &  & \textbf{486}     & \underline{\textbf{0}} & 238             &  & \textbf{486}     & \underline{\textbf{0}} & 19             &  & \textbf{486}     & \underline{\textbf{0}} & 17              &  & \textbf{486}     & \underline{\textbf{0}} & 33             &  & \textbf{486}     & \underline{\textbf{0}} & 17              \\
             & 9         &  & 455       & \cite{golden2012generalized,salari2012integer,salari2015combining,venkatesh2019multi} &  & 377              & 20.73            & 3600             &  & 438              & 3.88             & 3600            &  & \textbf{455}     & \underline{\textbf{0}} & 30             &  & \textbf{455}     & \underline{\textbf{0}} & 27              &  & \textbf{455}     & \underline{\textbf{0}} & 28             &  & \textbf{455}     & \underline{\textbf{0}} & 29              \\
             & 11        &  & 444       & \cite{golden2012generalized,salari2012integer,salari2015combining,venkatesh2019multi} &  & 350              & 26.81            & 3600             &  & \textbf{444}     & \underline{\textbf{0}} & 203             &  & \textbf{444}     & \underline{\textbf{0}} & 32             &  & \textbf{444}     & \underline{\textbf{0}} & 138             &  & \textbf{444}     & \underline{\textbf{0}} & 37             &  & \textbf{444}     & \underline{\textbf{0}} & 197             \\ \\
             \newpage

kroA100      & 7         &  & 9674      & \cite{golden2012generalized,salari2012integer,salari2015combining,venkatesh2019multi} &  & 9177             & 5.42             & 3600             &  & \textbf{9674}    & \underline{\textbf{0}} & 15              &  & \textbf{9674}    & \underline{\textbf{0}} & 18             &  & \textbf{9674}    & \underline{\textbf{0}} & 27              &  & \textbf{9674}    & \underline{\textbf{0}} & 27             &  & \textbf{9674}    & \underline{\textbf{0}} & 30              \\
             & 9         &  & 9159      & \cite{golden2012generalized,salari2012integer,salari2015combining,venkatesh2019multi} &  & 7938             & 15.38            & 3600             &  & \textbf{9159}    & \underline{\textbf{0}} & 154             &  & \textbf{9159}    & \underline{\textbf{0}} & 28             &  & \textbf{9159}    & \underline{\textbf{0}} & 2040            &  & \textbf{9159}    & \underline{\textbf{0}} & 36             &  & \textbf{9159}    & \underline{\textbf{0}} & 2230            \\
             & 11        &  & 8901      & \cite{golden2012generalized,salari2012integer,salari2015combining,venkatesh2019multi} &  & 8593             & 3.59             & 3600             &  & 8608             & 3.40             & 3600            &  & \textbf{8901}    & \underline{\textbf{0}} & 45             &  & 8640             & 3.02             & 3600            &  & \textbf{8901}    & \underline{\textbf{0}} & 79             &  & 8801             & 1.14             & 3600            \\
             &           &  &           &                                                                                                        &  &                  &                  &                  &  &                  &                  &                 &  &                  &                  &                &  &                  &                  &                 &  &                  &                  &                &  &                  &                  &                 \\
kroB100      & 7         &  & 9537      & \cite{golden2012generalized,salari2012integer,salari2015combining,venkatesh2019multi} &  & -                & -                & 3600             &  & \textbf{9537}    & \underline{\textbf{0}} & 45              &  & \textbf{9537}    & \underline{\textbf{0}} & 22             &  & \textbf{9537}    & \underline{\textbf{0}} & 20              &  & \textbf{9537}    & \underline{\textbf{0}} & 26             &  & \textbf{9537}    & \underline{\textbf{0}} & 23              \\
             & 9         &  & 9240      & \cite{golden2012generalized,salari2012integer,salari2015combining,venkatesh2019multi} &  & 7678             & 20.34            & 3600             &  & \textbf{9240}    & \underline{\textbf{0}} & 363             &  & \textbf{9240}    & \underline{\textbf{0}} & 21             &  & \textbf{9240}    & \underline{\textbf{0}} & 21              &  & \textbf{9240}    & \underline{\textbf{0}} & 31             &  & \textbf{9240}    & \underline{\textbf{0}} & 28              \\
             & 11        &  & 8842      & \cite{golden2012generalized,salari2012integer,salari2015combining,venkatesh2019multi} &  & -                & -                & 3600             &  & \textbf{8842}    & \underline{\textbf{0}} & 141             &  & \textbf{8842}    & \underline{\textbf{0}} & 25             &  & \textbf{8842}    & \underline{\textbf{0}} & 29              &  & \textbf{8842}    & \underline{\textbf{0}} & 40             &  & \textbf{8842}    & \underline{\textbf{0}} & 36              \\
             &           &  &           &                                                                                                        &  &                  &                  &                  &  &                  &                  &                 &  &                  &                  &                &  &                  &                  &                 &  &                  &                  &                &  &                  &                  &                 \\
kroC100      & 7         &  & 9723      & \cite{golden2012generalized,salari2012integer,salari2015combining}                    &  & 8564             & 13.54            & 3600             &  & \textbf{9723}    & \underline{\textbf{0}} & 561             &  & \textbf{9723}    & \underline{\textbf{0}} & 107            &  & \textbf{9723}    & \underline{\textbf{0}} & 102             &  & \textbf{9723}    & \underline{\textbf{0}} & 67             &  & \textbf{9723}    & \underline{\textbf{0}} & 92              \\
             & 9         &  & 9171      & \cite{golden2012generalized,salari2012integer,salari2015combining,venkatesh2019multi} &  & 7663             & 19.68            & 3600             &  & 8920             & 2.81             & 3600            &  & \textbf{9171}    & \underline{\textbf{0}} & 45             &  & \textbf{9171}    & \underline{\textbf{0}} & 783             &  & \textbf{9171}    & \underline{\textbf{0}} & 123            &  & \textbf{9171}    & \underline{\textbf{0}} & 972             \\
             & 11        &  & 8632      & \cite{golden2012generalized,salari2012integer,salari2015combining,venkatesh2019multi} &  & 7590             & 13.73            & 3600             &  & \textbf{8632}    & \underline{\textbf{0}} & 254             &  & \textbf{8632}    & \underline{\textbf{0}} & 38             &  & \textbf{8632}    & \underline{\textbf{0}} & 820             &  & \textbf{8632}    & \underline{\textbf{0}} & 222            &  & \textbf{8632}    & \underline{\textbf{0}} & 870             \\
             &           &  &           &                                                                                                        &  &                  &                  &                  &  &                  &                  &                 &  &                  &                  &                &  &                  &                  &                 &  &                  &                  &                &  &                  &                  &                 \\
kroD100      & 7         &  & 9626      & \cite{golden2012generalized,salari2012integer,salari2015combining,venkatesh2019multi} &  & 8724             & 10.34            & 3600             &  & \textbf{9626}    & \underline{\textbf{0}} & 59              &  & \textbf{9626}    & \underline{\textbf{0}} & 17             &  & \textbf{9626}    & \underline{\textbf{0}} & 20              &  & \textbf{9626}    & \underline{\textbf{0}} & 34             &  & \textbf{9626}    & \underline{\textbf{0}} & 23              \\
             & 9         &  & 8885      & \cite{golden2012generalized,salari2012integer,salari2015combining,venkatesh2019multi} &  & -                & -                & 3600             &  & \textbf{8885}    & \underline{\textbf{0}} & 16              &  & \textbf{8885}    & \underline{\textbf{0}} & 22             &  & \textbf{8885}    & \underline{\textbf{0}} & 27              &  & \textbf{8885}    & \underline{\textbf{0}} & 62             &  & \textbf{8885}    & \underline{\textbf{0}} & 35              \\
             & 11        &  & 8725      & \cite{golden2012generalized,salari2012integer,salari2015combining,venkatesh2019multi} &  & -                & -                & 3600             &  & \textbf{8725}    & \underline{\textbf{0}} & 51              &  & \textbf{8725}    & \underline{\textbf{0}} & 35             &  & \textbf{8725}    & \underline{\textbf{0}} & 48              &  & \textbf{8725}    & \underline{\textbf{0}} & 63             &  & \textbf{8725}    & \underline{\textbf{0}} & 80              \\
             &           &  &           &                                                                                                        &  &                  &                  &                  &  &                  &                  &                 &  &                  &                  &                &  &                  &                  &                 &  &                  &                  &                &  &                  &                  &                 \\
kroE100      & 7         &  & 10150     & \cite{golden2012generalized,salari2012integer,salari2015combining,venkatesh2019multi} &  & 9274             & 9.44             & 3600             &  & \textbf{10150}   & \underline{\textbf{0}} & 520             &  & \textbf{10150}   & \underline{\textbf{0}} & 81             &  & \textbf{10150}   & \underline{\textbf{0}} & 42              &  & \textbf{10150}   & \underline{\textbf{0}} & 76             &  & \textbf{10150}   & \underline{\textbf{0}} & 32              \\
             & 9         &  & 8991      & \cite{golden2012generalized,salari2012integer}                                        &  & 8500             & 5.77             & 3600             &  & \textbf{8991}    & \underline{\textbf{0}} & 336             &  & \textbf{8991}    & \underline{\textbf{0}} & 31             &  & \textbf{8991}    & \underline{\textbf{0}} & 55              &  & \textbf{8991}    & \underline{\textbf{0}} & 28             &  & \textbf{8991}    & \underline{\textbf{0}} & 88              \\
             & 11        &  & 8450      & \cite{golden2012generalized,salari2012integer,salari2015combining,venkatesh2019multi} &  & 7739             & 9.19             & 3600             &  & \textbf{8450}    & \underline{\textbf{0}} & 237             &  & \textbf{8450}    & \underline{\textbf{0}} & 23             &  & \textbf{8450}    & \underline{\textbf{0}} & 261             &  & \textbf{8450}    & \underline{\textbf{0}} & 33             &  & \textbf{8450}    & \underline{\textbf{0}} & 193             \\
             &           &  &           &                                                                                                        &  &                  &                  &                  &  &                  &                  &                 &  &                  &                  &                &  &                  &                  &                 &  &                  &                  &                &  &                  &                  &                 \\
rd100        & 7         &  & 3461      & \cite{golden2012generalized,salari2012integer,salari2015combining,venkatesh2019multi} &  & 3094             & 11.88            & 3600             &  & \textbf{3461}    & \underline{\textbf{0}} & 119             &  & \textbf{3461}    & \underline{\textbf{0}} & 20             &  & \textbf{3461}    & \underline{\textbf{0}} & 20              &  & \textbf{3461}    & \underline{\textbf{0}} & 24             &  & \textbf{3461}    & \underline{\textbf{0}} & 22              \\
             & 9         &  & 3194      & \cite{golden2012generalized,salari2012integer,salari2015combining,venkatesh2019multi} &  & 2664             & 19.90            & 3600             &  & \textbf{3194}    & \underline{\textbf{0}} & 63              &  & \textbf{3194}    & \underline{\textbf{0}} & 18             &  & \textbf{3194}    & \underline{\textbf{0}} & 18              &  & \textbf{3194}    & \underline{\textbf{0}} & 24             &  & \textbf{3194}    & \underline{\textbf{0}} & 25              \\
             & 11        &  & 2922      & \cite{golden2012generalized,salari2012integer,salari2015combining,venkatesh2019multi} &  & 2648             & 10.33            & 3600             &  & \textbf{2922}    & \underline{\textbf{0}} & 28              &  & \textbf{2922}    & \underline{\textbf{0}} & 21             &  & \textbf{2922}    & \underline{\textbf{0}} & 20              &  & \textbf{2922}    & \underline{\textbf{0}} & 34             &  & \textbf{2922}    & \underline{\textbf{0}} & 27              \\ \hline
\textbf{Avg} & \textbf{} &  & \textbf{} &                                                                                                        &  & \textbf{7673.50} & \textbf{9.27}    & \textbf{2867.39} &  & \textbf{8310.08} & \textbf{0.28}    & \textbf{396.75} &  & \textbf{8325.67} & \textbf{0.00}    & \textbf{23.28} &  & \textbf{8318.42} & \textbf{0.08}    & \textbf{229.19} &  & \textbf{8325.67} & \textbf{0.00}    & \textbf{35.69} &  & \textbf{8322.89} & \textbf{0.03}    & \textbf{243.92} \\ \hline
\end{longtable}
\end{landscape}

\begin{landscape}
\begin{table}[]
\setlength{\tabcolsep}{2pt}
\centering
\scriptsize
\caption{Results of computacional experiments for the medium-size instances.}
\begin{tabular}{lllllllllllllllllllllllll}
\hline
             &           &           & \multicolumn{2}{l}{BestUB}                                                                                         &           & \multicolumn{3}{l}{$CSP$-$I$}                           &           & \multicolumn{3}{l}{$CSP$-$I\&F_{vp}$}                &           & \multicolumn{3}{l}{$CSP$-$I\&F_{h}$}                    &           & \multicolumn{3}{l}{$CSP$-$I\&F_{vp}$-$X$}               &           & \multicolumn{3}{l}{$CSP$-$I\&F_{h}$-$X$}               \\ \cline{1-2} \cline{4-5} \cline{7-9} \cline{11-13} \cline{15-17} \cline{19-21} \cline{23-25} 
Instance     & NC        &           & UB        & Reference(s)                                                                                           &           & LB                & Gap          & Time             &           & LB                & Gap          & Time             &           & LB                & Gap          & Time             &           & LB                & Gap          & Time             &           & LB                & Gap          & Time            \\ \cline{1-2} \cline{4-5} \cline{7-9} \cline{11-13} \cline{15-17} \cline{19-21} \cline{23-25} 
kroA150      & 7         &           & 11423     & \cite{golden2012generalized,salari2012integer,salari2015combining,venkatesh2019multi} &           & 10658             & 7.18             & 3600             &           & \textbf{11423}    & \underline{\textbf{0}} & 174              &           & \textbf{11423}    & \underline{\textbf{0}} & 137              &           & \textbf{11423}    & \underline{\textbf{0}} & 147              &           & \textbf{11423}    & \underline{\textbf{0}} & 90              \\
             & 9         &           & 10056     & \cite{golden2012generalized,salari2012integer,salari2015combining,venkatesh2019multi} &           & \textbf{10056}    & \underline{\textbf{0}} & 147              &           & \textbf{10056}    & \underline{\textbf{0}} & 84               &           & \textbf{10056}    & \underline{\textbf{0}} & 85               &           & \textbf{10056}    & \underline{\textbf{0}} & 92               &           & \textbf{10056}    & \underline{\textbf{0}} & 122             \\
             & 11        &           & 9439      & \cite{golden2012generalized,salari2012integer,salari2015combining,venkatesh2019multi} &           & 9240              & 2.15             & 3600             &           & \textbf{9439}     & \underline{\textbf{0}} & 95               &           & \textbf{9439}     & \underline{\textbf{0}} & 67               &           & \textbf{9439}     & \underline{\textbf{0}} & 243              &           & \textbf{9439}     & \underline{\textbf{0}} & 91              \\
             &           &           &           &                                                                                                        &           &                   &                  &                  &           &                   &                  &                  &           &                   &                  &                  &           &                   &                  &                  &           &                   &                  &                 \\
kroB150      & 7         &           & 11457     & \cite{golden2012generalized,salari2012integer,salari2015combining,venkatesh2019multi} &           & 10663             & 7.45             & 3600             &           & \textbf{11457}    & \underline{\textbf{0}} & 334              &           & \textbf{11457}    & \underline{\textbf{0}} & 116              &           & \textbf{11457}    & \underline{\textbf{0}} & 113              &           & \textbf{11457}    & \underline{\textbf{0}} & 81              \\
             & 9         &           & 10121     & \cite{golden2012generalized,salari2012integer,salari2015combining,venkatesh2019multi} &           & 9951              & 1.71             & 3600             &           & \textbf{10121}    & \underline{\textbf{0}} & 280              &           & \textbf{10121}    & \underline{\textbf{0}} & 130              &           & \textbf{10121}    & \underline{\textbf{0}} & 145              &           & \textbf{10121}             & \underline{\textbf{0}}             & 112             \\
             & 11        &           & 9611      & \cite{golden2012generalized,salari2012integer,salari2015combining,venkatesh2019multi} &           & \textbf{9611}     & \underline{\textbf{0}} & 902              &           & \textbf{9611}     & \underline{\textbf{0}} & 849              &           & \textbf{9611}     & \underline{\textbf{0}} & 429              &           & \textbf{9611}     & \underline{\textbf{0}} & 947              &           & \textbf{9611}     & \underline{\textbf{0}} & 282             \\
             &           &           &           &                                                                                                        &           &                   &                  &                  &           &                   &                  &                  &           &                   &                  &                  &           &                   &                  &                  &           &                   &                  &                 \\
kroA200      & 7         &           & 13285     & \cite{golden2012generalized,salari2012integer}                                        &           & 11660             & 13.94            & 3600             &           & 12611             & 5.34             & 3600             &           & 12955             & 2.55             & 3600             &           & 12697             & 4.63             & 3600             &           & \textbf{13108}    & 1.35             & 3600            \\
             & 9         &           & 11708     & \cite{golden2012generalized,salari2012integer,venkatesh2019multi}                     &           & 10327             & 13.37            & 3600             &           & 11094             & 5.53             & 3600             &           & \textbf{11708}    & \underline{\textbf{0}} & 2252             &           & 11537             & 1.48             & 3600             &           & \textbf{11708}    & \underline{\textbf{0}} & 1008            \\
             & 11        &           & 10748     & \cite{golden2012generalized,salari2012integer}                                        &           & 9508              & 13.04            & 3600             &           & 10342             & 3.93             & 3600             &           & \textbf{10748}    & \underline{\textbf{0}} & 1044             &           & \textbf{10748}    & \underline{\textbf{0}} & 3582             &           & \textbf{10748}    & \underline{\textbf{0}} & 648             \\
             &           &           &           &                                                                                                        &           &                   &                  &                  &           &                   &                  &                  &           &                   &                  &                  &           &                   &                  &                  &           &                   &                  &                 \\
kroB200      & 7         &           & 13051     & \cite{salari2012integer,salari2015combining,venkatesh2019multi}                       &           & 12260             & 6.45             & 3600             &           & 12462             & 4.73             & 3600             &           & 12904             & 1.14             & 3600             &           & 12697             & 2.79             & 3600             &           & \textbf{13051}    & \underline{\textbf{0}} & 1487            \\
             & 9         &           & 11864     & \cite{salari2012integer,salari2015combining,venkatesh2019multi}                       &           & 11209             & 5.84             & 3600             &           & 11379             & 4.26             & 3600             &           & \textbf{11864}    & \underline{\textbf{0}} & 2281             &           & 11695             & 1.45             & 3600             &           & \textbf{11864}    & \underline{\textbf{0}} & 1242            \\
             & 11        &           & 10644     & \cite{salari2012integer,salari2015combining,venkatesh2019multi}                       &           & 10405             & 2.30             & 3600             &           & \textbf{10644}    & \underline{\textbf{0}} & 800              &           & \textbf{10644}    & \underline{\textbf{0}} & 907              &           & \textbf{10644}    & \underline{\textbf{0}} & 514              &           & \textbf{10644}    & \underline{\textbf{0}} & 938             \\ \hline
\textbf{Avg} & \textbf{} & \textbf{} & \textbf{} & \textbf{}                                                                                              & \textbf{} & \textbf{10462.33} & \textbf{6.12}    & \textbf{3087.42} & \textbf{} & \textbf{10886.58} & \textbf{1.98}    & \textbf{1718.00} & \textbf{} & \textbf{11077.50} & \textbf{0.31}    & \textbf{1220.67} & \textbf{} & \textbf{11010.50} & \textbf{0.86}    & \textbf{1681.92} & \textbf{} & \textbf{11102.42} & \textbf{0.11}    & \textbf{808.42} \\ \hline
\end{tabular}
\label{tab:mediumInstances}
\end{table}
\end{landscape}

\restoregeometry

For small-size instances, there were previously known lower bounds for $32$ out of $36$ instances, obtained by $SRS$, from which optimal solutions were proven for $9$ instances. The proposed branch-and-cut framework, on the other hand, obtained lower bounds for all instances. More importantly, the framework proved optimality for all $36$ small instances. All branch-and-cut methodologies outperformed $SRS$ with respect to optimality gap, and they were fairly robust among themselves; the worst performing ($CSP$-$I$) obtained an average $0.28\%$ optimality gap, while the best performing ($CSP$-$I\&F_{vp}$ and $CSP$-$I\&F_{vp}$-$X$) with zero optimality gap, shows the exact separation prevails over the heuristic separation of fractional solutions for small instances.


With respect to medium-size instances, no lower bound was known for any of the $12$ instances in the literature. The branch-and-cut framework obtained the first lower bounds for all these instances. Furthermore, optimality was proven for all instances except one ($kroA200$-$7$), which remains with an optimality gap of $1.35\%$. The performance among the branch-and-cut methodologies varied more significantly this time. The best-performing methodology was $CSP$-$I\&F_{h}$-$X$, with an average  gap of $0.11\%$. Now, the heuristic separation overcomes the exact separation, mainly due to the reduction in the computational effort. The worst-performing methodology ($CSP$-$I$) obtained an average gap of $6.12\%$, showing that by using only integral cuts performs poorly for more challenging instances.

The effect of the CI inequalities \eqref{eq:newInequalityCSP} in the performance of the methodologies was also examined. Comparing $CSP$-$I\&F_{vp}$ and $CSP$-$I\&F_{vp}$-$X$, their average gaps were both zero for small instances and reduced from $1.98\%$ to $0.86\%$ for medium instances. Moreover, comparing $CSP$-$I\&F_{h}$ and $CSP$-$I\&F_{h}$-$X$, their average gaps reduced from $0.08\%$ to $0.03\%$ for small instances and reduced from $0.31\%$ to $0.11\%$ for medium instances. Therefore, the CI inequalities are confirmed to have a significant impact on reducing the optimality gaps. 

Previously, from $48$ small and medium-size CSP instances, only $9$ optimal solutions were known. These computational results have shown that the branch-and-cut framework, by borrowing meaningful valid inequalities from GTSP and proposing  new valid inequalities for CSP, was able to obtain optimal solutions for all instances except one, thus $38$ instances were proven optimal for the first time.

\section{Final Remarks} \label{sec:conclusions}

The proposed branch-and-cut framework for the CSP uses existing valid inequalities for the GTSP, by Fischetti et al. \cite{fischetti1997branch}, and a new family of valid inequalities, \textit{CI inequalities}, to improve on the state-of-the-art exact methodology for the CSP. Exact and heuristic separation routines for integer and fractional solutions are investigated.


The branch-and-cut framework is composed of five methodologies using distinct families of inequalities and separation routines. Computational experiments conducted on a benchmark of $48$ instances from literature delves into the effectiveness of the framework. 
The overall results show unequivocally the branch-and-cut methodologies outperforming the best known exact methodology from literature and unveiling $38$ new optimal solutions.
The experiments also show that the CI inequalities had a major role in the performance of the methodologies.

The ideas presented in this work can support the exact solution of many possible developments of the CSP. Future works may consider, for example, CSP with multiple vehicles, capacity constraints, time constraints, green vehicles, uncertainty on the covering neighborhood, and other generalizations of the CSP which better approximate practical routing problems. The new family of valid inequalities proposed in this work should be considered on the exact solution for any of these generalizations.

\section*{Acknowledgments}

This work was supported by CAPES, CNPq, and Fapesp (grants 140960/2017-1, 314384/2018-9, 435520/2018-0, 2015/11937-9).

\bibliographystyle{unsrt}  
\bibliography{references}  

\end{document}